\documentclass[sigconf,nonacm]{acmart}

\usepackage[english]{babel}
\usepackage{blindtext}
\usepackage{multirow}
\usepackage[dvipsnames]{xcolor}
\usepackage{colortbl}
\usepackage{adjustbox}
\usepackage{subfigure}
\usepackage{graphicx}
\graphicspath{ {figures/} }
\usepackage{comment}
\usepackage{tabularx}
\usepackage{hyphenat}

\usepackage{diagbox}

\newcommand{\ie}{{\em i.e., }}
\newcommand{\eg}{{\em e.g., }}

\newcommand{\mycomment}[1]{}

\newcommand\colorlaunch[1]{\textcolor{RedOrange}{#1}}
\newcommand\coloridle[1]{\textcolor{RoyalBlue}{#1}}
\newcommand\colorpassive[1]{\textcolor{Purple}{#1}}
\newcommand\coloractive[1]{\textcolor{ForestGreen}{#1}}

\begin{document}
\title[Games Are Not Equal: Classifying Cloud Gaming Contexts for Effective User Experience Measurement]{Games Are Not Equal: Classifying Cloud Gaming Contexts\\ for Effective User Experience Measurement}\thanks{This is the preprint version of our paper \cite{YWangIMC25} accepted at ACM IMC 2025.}

\author{Yifan Wang}
\orcid{0009-0009-3472-9821}
\affiliation{%
	\institution{University of New South Wales}
        \city{Sydney}
        \state{NSW}
	\country{Australia}
}
\email{wangyifan.frank@student.unsw.edu.au}

\author{Minzhao Lyu}
\orcid{0000-0001-8677-248X}
\affiliation{%
	\institution{University of New South Wales}
        \city{Sydney}
        \state{NSW}
	\country{Australia}
}
\email{minzhao.lyu@unsw.edu.au}\thanks{Corresponding author: Minzhao Lyu (minzhao.lyu@unsw.edu.au)}

\author{Vijay Sivaraman}
\orcid{0000-0001-7985-6765}
\affiliation{%
	\institution{University of New South Wales}
        \city{Sydney}
        \state{NSW}
	\country{Australia}
}
\email{vijay@unsw.edu.au}



\begin{abstract}
To tap into the growing market of cloud gaming, whereby game graphics is rendered in the cloud and streamed back to the user as a video feed, network operators are creating monetizable assurance services that dynamically provision network resources. However, without accurately measuring cloud gaming user experience, they cannot assess the effectiveness of their provisioning methods. Basic measures such as bandwidth and frame rate by themselves do not suffice, and can only be interpreted in the context of the game played and the player activity within the game.
This paper equips the network operator with a method to obtain a real-time measure of cloud gaming experience by analyzing network traffic, including contextual factors such as the game title and player activity stage. Our method is able to classify the game title within the first five seconds of game launch, and continuously assess the player activity stage as being active, passive, or idle. We deploy it in an ISP hosting NVIDIA cloud gaming servers for the region. We provide insights from hundreds of thousands of cloud game streaming sessions over a three-month period into the dependence of bandwidth consumption and experience level on the gameplay contexts. 
\end{abstract}

\maketitle

\section{Introduction} \label{sec:intro} 

Cloud gaming has gained significant popularity as an affordable and accessible alternative to traditional gaming by shifting intensive real-time graphics rendering and gaming computation from client hardware to cloud GPU servers. Users on high-speed networks can play the latest game releases without high-end graphics computing hardware. Driven by the increasing demand for high-quality gaming, the maturity of cloud GPU infrastructure, and the increase in the capacity of carrier networks \cite{ericsson_5g_2023}, the cloud gaming industry, led by tech giants such as NVIDIA \cite{nvidia_geforcenow}, Microsoft \cite{microsoft_xcloud}, Amazon \cite{amazon_luna} and Sony \cite{sony_ps5cloud}, is expected to soar to US\$143bn by 2032 \cite{yahoo2024}.

Network operators are experimenting with ways to monetize cloud gaming, by providing network assurance for the service. Unlike traditional online games that only require a few hundred Kbps of bandwidth from the network \cite{madanapalli_know_2022}, cloud games need much higher bandwidth of tens of Mbps; simultaneously, unlike traditional video streaming that can buffer seconds or even minutes of content, cloud game streaming experience is impacted by sub-second glitches in latency or packet drops \cite{lyu_do_2024}. This highly demanding nature of cloud gaming in terms of both bandwidth and latency/loss renders it amenable for a premium (monetizable) assurance service, such as via a 5G slice dedicated to cloud gaming \cite{vodafone_cloudgaming_5g}, or via dynamic provisioning \eg using the Quality-on-Demand API being standardized by the CAMARA group in the TM Forum \cite{camara_qod_api}. As an example, Telefonica announced in 2023 that it was conducting such a trial with the Blacknut Cloud Gaming service \cite{telefonica_blacknut_qod}, followed more recently by Singtel's collaboration with Tencent Games to launch a specialized 5G slice service for cloud game streaming \cite{singtel_tencent_2025} in 2025.

\begin{figure*}[t!]
	\centering
	\subfigure[The gameplay activity follows a divided \textbf{spectate-and-play} pattern in this shooter (CS:GO) cloud gaming session.]{
		\includegraphics[width=0.485\textwidth]{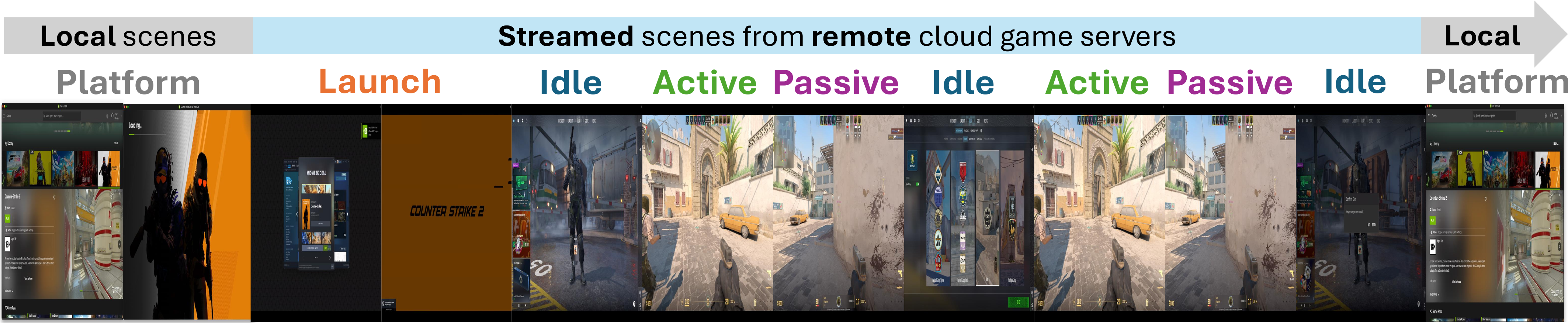}
		\label{fig:fortnite_process}
	}
	\subfigure[The gameplay activity follows a nearly \textbf{continuous-play} pattern in this role-playing (Cyberpunk 2077) cloud gaming session.]{
		\includegraphics[width=0.485\textwidth]{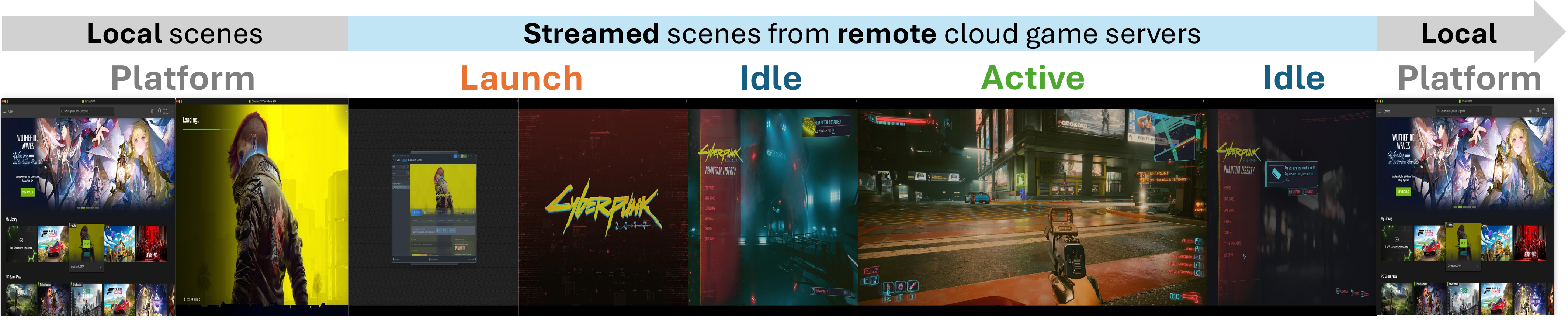}
		\label{fig:cyberpunk_process}
	}
   \vspace{-5mm}
	\caption{Examples of the two types of gameplay activity patterns including (a) a shooter game session where player activities follow a spectate-and-play pattern and (b) a role-playing game session where player activities follow a continuous-play pattern.}
	\label{fig:main}
    \vspace{-3mm}
\end{figure*}

For a network operator to assure the gaming experience, they have to be able to measure the gaming experience accurately. This is not as trivial as measuring the throughput, latency, and loss of the game video stream, because these vary depending on the title/genre of the game \cite{bhuyan_end_2022,sabet_delay_2020}, as well as the activity stage of the player within the game \cite{carrascosa_cloud_2022,moll_network_2018}, namely whether they are active (\eg shooting), passive (\eg spectating), or idle (\eg between rounds). In the absence of this context, drops in bandwidth or streaming frame rate may wrongly be associated with poor experience, when in reality the user is engaged in a less intensive role-playing game or a strategy card game, or spectating others in a shooting/sports game.
For example, players in a First Person Shooter (FPS) game often actively engage in gameplay when a competitive match starts with stringent requirements on high streaming frame rate, and have relatively lower network requirements during passive activity stage such as matchmaking and observation. As another example, players of a card game are likely to have nearly static gaming scenarios with low expectations on streaming frame rate and bandwidth during the entire gameplay.

In this paper, we develop a novel method to help network operators measure the \textbf{context} of cloud gaming sessions, specifically the title/genre of the game and the player activity stage (\ie active, passive or idle) within the gameplay. This can be combined with prior works that have developed traffic analysis methods to classify cloud gaming sessions, identify user platforms, and measure objective Quality-of-Experience (QoE) metrics from network Quality-of-Service (QoS) parameters \cite{lyu_network_2024,carrascosa_cloud_2022,domenico_network_2021}, as well as studies that have benchmarked the degradation of game streaming quality due to network conditions \cite{lyu_do_2024,marchal_analysis_2023,ky_ml_2023,graff_analysis_2021,aumont_dissecting_2021}. With our three contributions listed below, this paper fills a gap by providing network operators with an accurate measure of the \textbf{effective} user experience that takes into account the context of game title/genre and player activity stage.

Our \textbf{first} contribution (\S\ref{sec:dissect}) systematically identifies the unique network traffic characteristics of cloud game streaming sessions in various gameplay contexts. We collect and study traffic traces of over 500 cloud gaming sessions across 13 most popular games spanning 5 genres and diverse game streaming settings. The labeled traffic dataset will be shared with the community. We reveal the unique statistical distributions in packet sizes and timings during the initial launch stage of each cloud game title/genre, as well as the relative volumetric patterns within each streaming session that relate to the player activity stages in the game.

In the \textbf{second} contribution (\S\ref{sec:detect}), we develop and evaluate a real-time network traffic analysis method to classify the cloud game title within the first five seconds of game launch, and continuously measure the player activity stage thereafter. Game title classification is achieved using pre-trained machine learning models on packet sizes and timing attributes from the first five seconds, whereas the player activity stage is continuously assessed from bidirectional volumetric profiles. When the fine-grained game title cannot be confidently inferred, our method infers the coarse-grained game genre (\ie gameplay activity pattern) from the transition behaviors between player activity stages.

Our \textbf{third} contribution (\S\ref{sec:insights}) deploys our cloud gameplay context classification into a live network operated by our partner Internet service provider (ISP) that hosts NVIDIA's GeForce NOW cloud gaming servers for our region. We collect data over a three-month period, spanning hundreds of thousands of cloud gaming hours. We validate our game title/genre classification method determined in real-time against ground truth from the game server logs produced offline after the conclusion of each gameplay session. We then highlight how bandwidth demand varies depending on the game title/genre, as well as the player activity stage within the game. Lastly, we show that by providing appropriate game contexts, our method can help network operators avoid mislabeling experience drops that are attributable to the game title/genre or player activity rather than poor network conditions, thereby helping them accurately gauge and assure cloud gaming experience over their network.

\section{Preliminary of User Experience Contexts in Cloud Gaming}\label{sec:background}

In this section, we provide preliminaries on contextual factors of cloud gameplay that can affect user expectations on their perceived game streaming quality, starting from an intuitive understanding of typical player activity patterns during cloud gameplay sessions (\S\ref{sec:cg_process}), to game contexts (\ie titles/genres) that exhibit unique activity patterns (\S\ref{sec:cg_contexts}).

\subsection{Player Activity Stage and Gameplay Activity Pattern}\label{sec:cg_process}
Prior works \cite{lyu_network_2024,marchal_analysis_2023} have revealed that players who access commercial cloud gaming services first connect to the respective cloud platform for administrative services, game selection, and allocation of cloud game servers before actual gameplay. However, diversified user activities during cloud gameplay sessions that can lead to different user expectations on their perceived streaming quality, network traffic characteristics, and network demands to be satisfied by network operators have not yet been studied. Therefore, we now zoom into the gameplay session to understand the various types of user activities categorized into three levels of \textbf{\textit{player activity stages}}, namely \textbf{active}, \textbf{passive} and \textbf{idle}, which can lead to various expectations on the streaming quality.

As visually shown in Fig.~\ref{fig:main}, we use two representative gameplay sessions on NVIDIA's GeForce NOW platform to discuss the \textbf{\textit{gameplay activity patterns}} each with a different profile of player activity stages. Our discussed insights are generalizable to other commercial platforms.

The first example in Fig.~\ref{fig:fortnite_process} shows a CS:GO shooter game session where players' gameplay activities can be divided into multiple \textbf{spectate-and-play} slots. 
In this shooter gameplay, after the game has been selected on the cloud gaming platform, the player first waits for the game to launch with a transition scene and then stays in the lobby until a match starts. During these periods, the player maintains an almost idle level of gameplay activity. As will be discussed later in \S\ref{sec:dissect}, such idle activities also exhibit low-profile traffic characteristics.
Once the player starts a match, the gameplay activity becomes very active, including both high-frequency graphics refreshment and user input updates. During this shooter game match, the player sometimes gets eliminated by opponents and waits to respawn, becoming passive in terms of gameplay activity by spectating with decent graphics refresh frequency but very limited user interaction. The player goes back to the idle activity stage in the lobby before starting the second match. In realistic gameplay sessions, the spectate-and-play pattern often occurs multiple times (twice in this example) before the session ends.

\begin{table}[t!]
	\caption{Thirteen popular cloud game titles played on NVIDIA's GeForce NOW platform in our geography, with their game genre, gameplay activity pattern, and popularity by their fraction of total playtime.}
	\centering
	\resizebox{\columnwidth}{!}{
		\begin{tabular}{|l|l|l|l|}
			\hline
			\rowcolor[HTML]{C0C0C0} 
			\textbf{Game title}                    & \textbf{Game genre}                                  & \textbf{Activity pattern}                    & \textbf{Popularity} \\ \hline
			Fortnite                 & Shooter                          & Spectate-and-play                         & 37.80\%    \\ \hline
			Genshin Impact                  & Role-playing                             & Continuous-play                       & 20.10\%    \\ \hline
			Baldur's Gate 3           & Role-playing                          & Continuous-play                        & 3.30\%     \\ \hline
			R6: Siege                & Shooter                                    & Spectate-and-play               & 1.24\%     \\ \hline
			Honkai: Star Rail      & Role-playing                          & Continuous-play                       & 1.16\%     \\ \hline
			Destiny 2                & Shooter                                    & Spectate-and-play                  & 1.15\%     \\ \hline
			Call of Duty             & Shooter                                    & Spectate-and-play             & 0.97\%     \\ \hline
			Cyberpunk 2077           & Role-playing                             & Continuous-play             & 0.84\%     \\ \hline
			Overwatch 2              & Shooter                                    & Spectate-and-play             & 0.74\%     \\ \hline
			Rocket League            & Sports                                 & Spectate-and-play                  & 0.64\%     \\ \hline
			CS:GO/CS2                 & Shooter                                    & Spectate-and-play                        & 0.61\%     \\ \hline
			Dota 2                   & MOBA                                   & Spectate-and-play                        & 0.55\%     \\ \hline
			Hearthstone     & Card & Spectate-and-play                  & 0.04\%     \\ \hline
	\end{tabular}}
    \vspace{-3mm}
	\label{tab:game_selection}
\end{table}

The second representative gameplay is shown in Fig.~\ref{fig:cyberpunk_process}, where the user plays Cyberpunk 2077, a role-playing game that follows a \textbf{continuous-play} activity pattern. 
For this type of gameplay activity pattern, players often engage in active gameplay activity in a continuous manner for a relatively long period, except at the beginning (\eg login and character selection) and end of the session.
Under such game mechanism, during the active gameplay stage, the player explores the game world without interruption and completes multiple missions, with only occasional short occurrences of passive or idle stages (\eg static dialogue or transition animations) featuring infrequent graphics refresh and user motion updates.
As will be discussed in \S\ref{sec:dissect}, gameplay following the continuous-play player activity pattern has its traffic profile at a constantly high level with small fluctuations.

\subsection{Cloud Game Titles and Genres}\label{sec:cg_contexts}
Having discussed the three categories of player activity stages (\ie idle, passive, and active) observable within the two types of gameplay activity patterns (\ie spectate-and-play and continuous-play), each of which can result in unique expectations on game streaming quality such as frame rate and, therefore, network bandwidth demands, we now discuss the gameplay contexts including game titles and genres that inherently lead to these distinct gameplay activity patterns.

We have collected an exhaustive list of cloud game catalog from our partnered network operator hosting NVIDIA's GeForce NOW cloud gaming servers in our geography. While we cannot disclose the full list as requested by our industry partner, we present in Table~\ref{tab:game_selection} the top 13 game titles that contribute to over 69\% of total gameplay time during a half-year period encompassing millions of gameplay sessions, with their genres defined by the gaming community, popularity in percentage of gameplay time, and their respective gameplay activity patterns observed in our study listed.

\begin{figure}[t!]
	\includegraphics[width=\columnwidth]{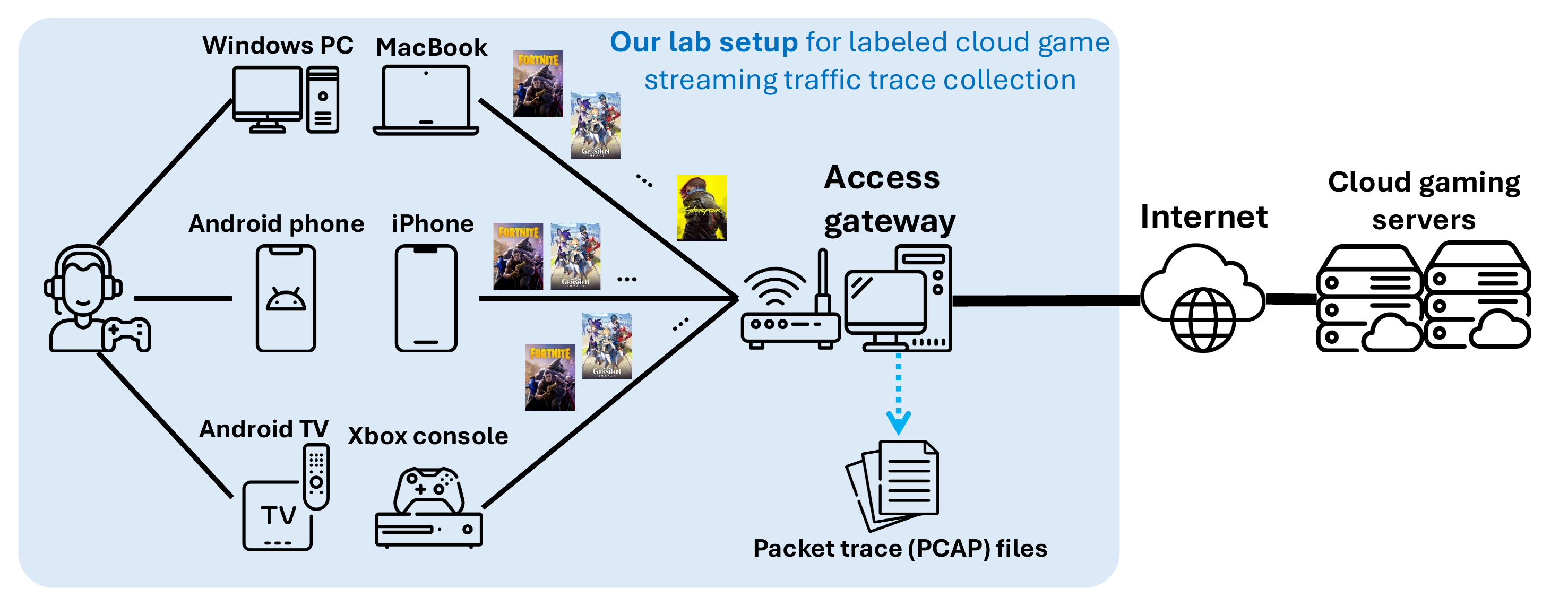}
	\vspace{-3mm}
	\caption{Lab experiment setup for collecting traffic traces of cloud gaming sessions.}
	\label{fig:lab_setup}
\end{figure}

\begin{table}[t!]
	\caption{Our lab traffic capture dataset of NVIDIA's GeForce NOW cloud gameplay sessions across the thirteen popular game titles on a diverse profiles of user configuration.}
	\centering
	\resizebox{\columnwidth}{!}{
		\begin{tabular}{|l|l|l|l|l|l|}
			\hline
			\rowcolor[HTML]{C0C0C0} 
			\textbf{Device}        & \textbf{OS}           & \textbf{Software} & \textbf{Streaming settings} & \textbf{\#Sessions} & \textbf{Playtime} \\ \hline
			&                           & Native app        & UHD-SD; 30-120 fps               & 89                  & 10.9 hours              \\ \cline{3-6} 
			& \multirow{-2}{*}{Windows} & Browser           & QHD-SD; 30-120 fps               & 60                  & 6.8 hours                   \\ \cline{2-6} 
			&                           & Native app        & UHD-SD; 30-120 fps               & 76                  & 10.5 hours                  \\ \cline{3-6} 
			\multirow{-4}{*}{PC}     & \multirow{-2}{*}{macOS}   & Browser           & QHD-SD; 30-120 fps               & 61                  & 7.7 hours                   \\ \hline
			& Android                   & Native app        & QHD-FHD; 30-120 fps                  & 73                  & 9.1 hours                   \\ \cline{2-6} 
			\multirow{-2}{*}{Mobile} & iOS                       & Browser           & FHD-SD; 30-120 fps                    & 70                  & 8.8 hours                   \\ \hline
			TV                 & AndroidTV                & Native app        & FHD-SD; 30-120 fps                    & 48                  & 6.1 hours                   \\ \hline
			Console           & Xbox                      & Browser           & FHD-SD; 30-120 fps                    & 54                  & 7.1 hours                   \\ \hline
		\end{tabular}
	}
	\label{tab:dataset}
	\vspace{-3mm}
\end{table}

The top 13 game titles belong to five game genres as defined by the gaming community based on their content, including shooter, role-playing, sports, MOBA, and card games. The two examples just discussed in \S\ref{sec:cg_process} are both in the list. By investigating into all the thirteen game titles on the GeForce NOW platform as well as popular games on three other major cloud gaming platforms, we validated that the gameplay activity patterns (\ie spectate-and-play and continuous-play) as articulated in \S\ref{sec:cg_process} are directly correlated with their game title and genre. For example, all the six shooter games in the list have their gameplay activities following the spectate-and-play pattern with a repeating combination of idle, active, and/or passive player activity stages, so do MOBA and card games. Player activities in all cloud gameplay sessions of role-playing games are constantly active and thus follow the continuous-play pattern. 

\begin{figure*}[t!]
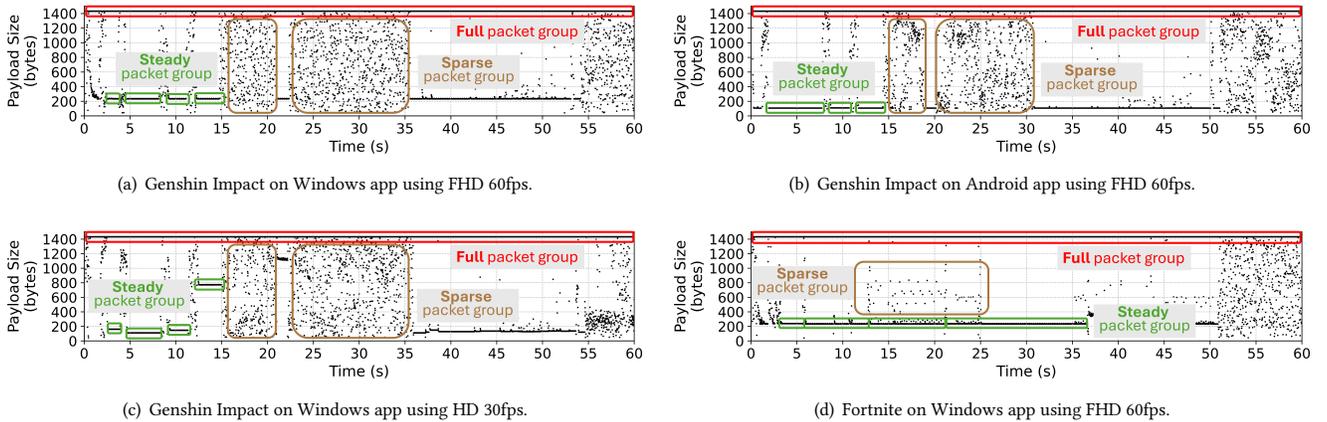

	\mbox{
		\subfigure[Genshin Impact on Windows app using FHD 60fps.]{
			\includegraphics[width=0.485\textwidth]{figures/packet_genshin_win_compressed.pdf}
			\label{fig:packet_genshin_win}
		}
		\subfigure[Genshin Impact on Android app using FHD 60fps.]{
			\includegraphics[width=0.485\textwidth]{figures/packet_genshin_android_compressed.pdf}
			\label{fig:packet_genshin_android}
		}
	}
	\mbox{
		\subfigure[Genshin Impact on Windows app using HD 30fps.]{
			\includegraphics[width=0.485\textwidth]{figures/packet_genshin_win_hd_compressed.pdf}
			\label{fig:packet_genshin_win_hd}
		}
		\subfigure[Fortnite on Windows app using FHD 60fps.]{
			\includegraphics[width=0.485\textwidth]{figures/packet_fortnite_win_compressed.pdf}
			\label{fig:packet_fortnite_win}
		}
	}
   \vspace{-3mm}
	\caption{Time-series scatter plots for the downstream packet payload sizes arriving in the first 60 seconds (covering the initial game launching stage) of four representative cloud game streaming sessions, where the packets can be categorized into the full, steady, or sparse groups.}
    \label{fig:packet_patterns}
\end{figure*}

As measured in \cite{lyu_do_2024,bhuyan_end_2022}, cloud gaming servers render gaming graphics that are streamed back to players' devices with different quality and encoding granularity depending on the game titles. Soon in \S\ref{sec:dissect}, we will discuss that the network traffic characteristics (\eg bandwidth usage) vary across not only game titles and gameplay activity patterns, but also player activity stages within each session, leading to diversified user expectations on streaming quality and thus network demands of the particular session. Therefore, knowing the cloud game contexts (\ie game titles/genres and player activity stages) is essential for network operators to accurately understand the effective cloud gaming quality delivered over their network infrastructure to the subscribers, so that they can precisely troubleshoot those that are indeed impacted by poor network conditions for proactive network optimization.
\section{Traffic Characterization across Cloud Game Contexts}\label{sec:dissect}
In this section, using our ground-truth traffic traces of cloud gaming sessions labeled by game titles and in-game player activity stages (\S\ref{sec:dataset}), we discuss our findings into the unique group characteristics of packets during game launch across titles (\S\ref{sec:dissect_game_titles}) and the volumetric characteristics of streaming flows in the upstream and downstream directions across player activity stages (\S\ref{sec:dissect_engagement_status}).

\subsection{Dataset Collection and Overview}\label{sec:dataset}
To the best of our knowledge, existing public traffic trace datasets (\eg \cite{slivar_cgd_2022,slivar_game_2018,marchal_analysis_2023,carrascosa_cloud_2022}) on cloud gaming do not contain sufficient labels of the game contexts articulated in \S\ref{sec:background}, and/or not have comprehensive coverage of game streaming configurations. Therefore, to obtain a holistic understanding of the network traffic characteristics of cloud gameplay sessions with diversified configurations (\ie user devices, operating systems, software applications and streaming settings) across different types of cloud game contexts (\ie game titles, genres and player activity stages), we built a lab setup to collect ground-truth traffic capture (PCAP) files of cloud game streaming sessions, as visually shown in Fig.~\ref{fig:lab_setup}.

The setup consists of a diverse collection of user devices that support cloud game streaming services, including mobile phones, PCs, a gaming console and a smart TV. The devices are connected to our regional NVIDIA GeForce NOW cloud servers hosted by our partnered Internet service provider with network conditions of nearly 1 Gbps bandwidth, less than 10 ms latency and less than 0.1\% packet loss rate. They can also access cloud gaming services hosted by other platforms such as Microsoft's Xbox Cloud Gaming with nearly ideal network conditions. The Linux-based access gateway connecting user devices is configured with packet capturing capability through Wireshark and TCPdump software.

With a team comprising two of the authors and four student volunteers, we have collected labeled PCAP files for 531 cloud game streaming sessions with 67 hours of total playtime across the 13 popular game titles listed in Table~\ref{tab:game_selection}. 
The number of game streaming sessions and their total playtime of the collected traffic traces labeled by their device genres, operating systems, software types, and streaming settings are presented in Table~\ref{tab:dataset}.
To enable us to analyze traffic characteristics associated not only with per-session game titles but also with in-game player activity stages, each PCAP file that contains packet streams of an entire gameplay session is also labeled by the changing player activity stages (\ie idle, active, or passive) with timestamps.

To validate the generalizability of our insights to other commercial cloud gaming platforms, we have also collected labeled PCAP files for a representative range of gameplay sessions on three other major platforms including Xbox Cloud Gaming, Amazon Luna, and PS5 Cloud Streaming, in addition to the PCAPs for GeForce NOW. The dataset is about 800 GB in size and is shared with the Internet research community through our university cloud drive with the availability information provided in Appendix \S\ref{sec:data_availability}.

\begin{figure*}[t!]
	\centering
	\subfigure[Overwatch using HD resolution.]{
		\includegraphics[width=0.48\textwidth]{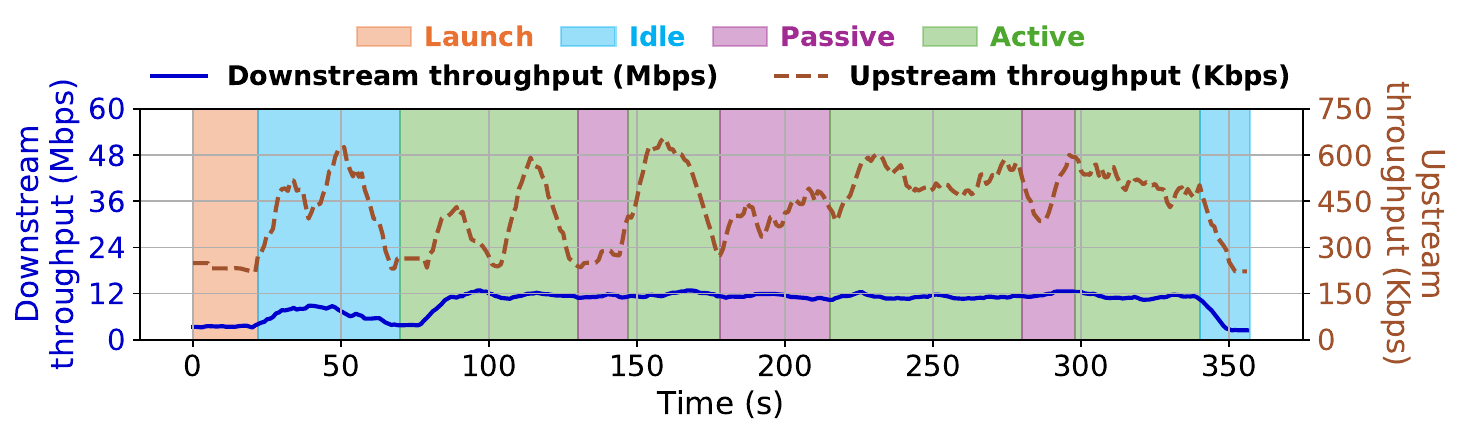}
		\label{fig:volumetric_overwatch_win_hd}
	}
	\subfigure[Overwatch using UHD resolution.]{
		\includegraphics[width=0.48\textwidth]{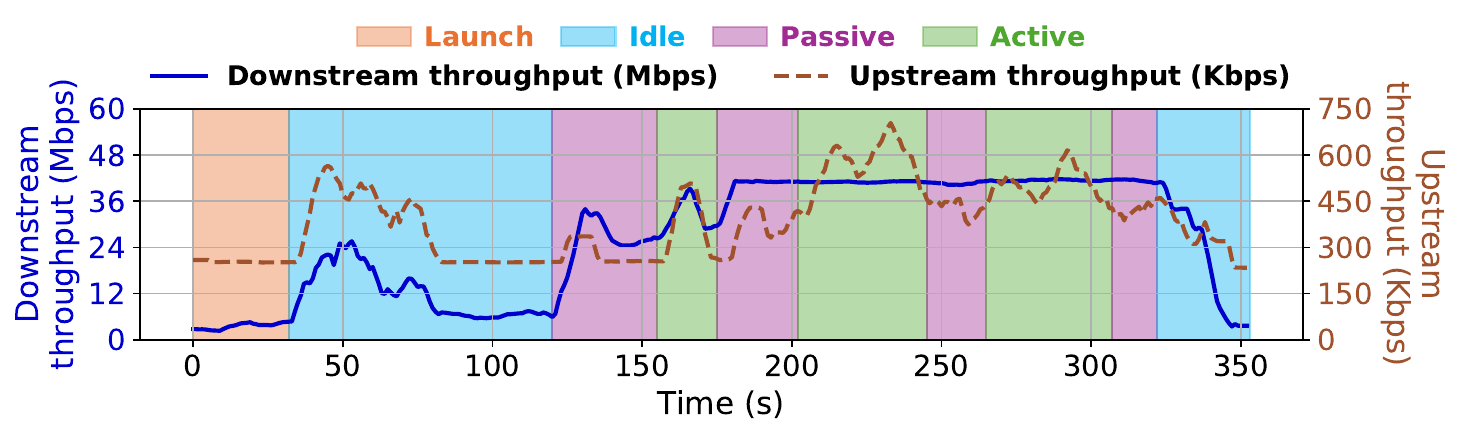}
		\label{fig:volumetric_overwatch_win_uhd}
	}
	\subfigure[CS:GO using UHD resolution.]{
		\includegraphics[width=0.48\textwidth]{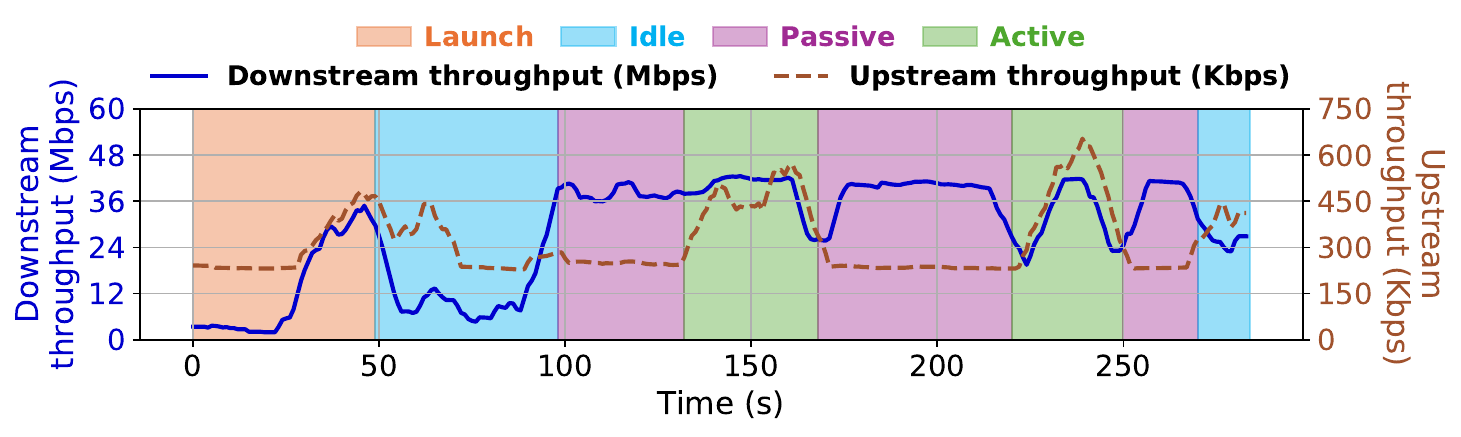}
		\label{fig:volumetric_csgo_win_uhd}
	}
	\subfigure[Cyberpunk 2077 using UHD resolution.]{
		\includegraphics[width=0.48\textwidth]{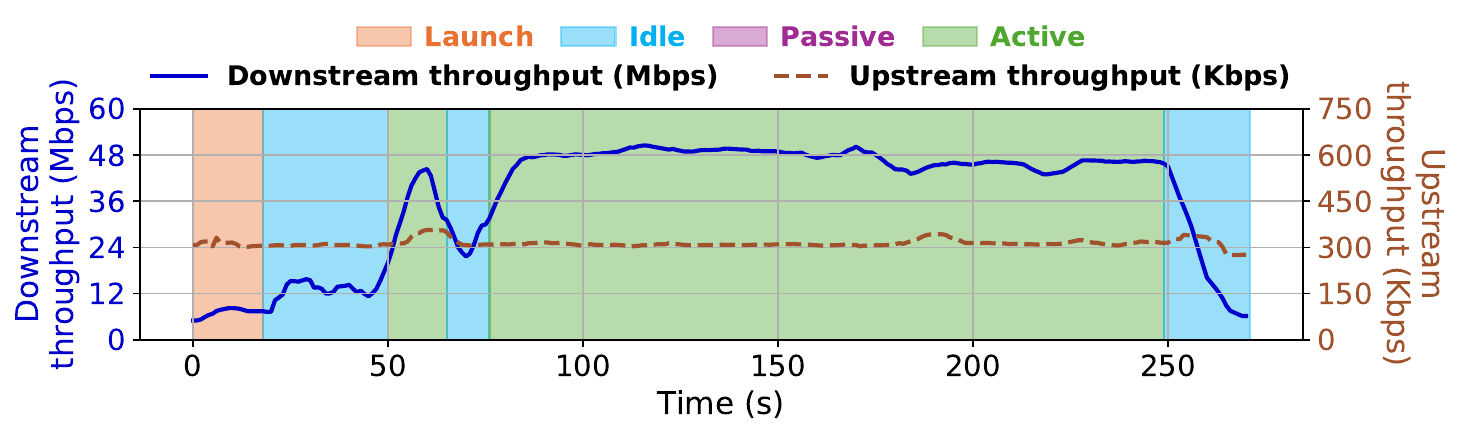}
		\label{fig:volumetric_cyberpunk_mac_uhd}
	}
   \vspace{-3mm}
	\caption{Time-series line chart showing the downstream throughput and upstream packet rate of game streaming flows in four representative cloud gaming sessions with background color-coded by player activity stages as \colorlaunch{launch}, \coloridle{idle}, \colorpassive{passive} and \coloractive{active}.}
	\label{fig:session_volumetric}
\end{figure*}

\subsection{Packet Group Profiles across Game Titles} \label{sec:dissect_game_titles}

During cloud gameplay, cloud servers stream the rendered gaming video and receive user inputs in real-time through standard Real-time Transport Protocol (RTP) flows \cite{nvidia_how_2023,lyu_network_2024,domenico_network_2021}.
Specifically, gaming graphics and audio are sent from cloud servers to the client devices via downstream RTP packets of the respective streaming flows, whereas player inputs such as mouse clicks, key strokes, mobile screen touches, and voice are carried in upstream RTP packets from the client devices to cloud servers. Prior works \cite{lyu_network_2024,domenico_network_2021} have also developed signatures to detect RTP flows for cloud game streaming using flow metadata, which are used as preliminary filtering conditions on the collected traffic traces for us to focus on the game streaming flows and investigate their behavioral profiles across game contexts.

As already illustrated in Fig.~\ref{fig:main}, before a player starts actively or passively engaging in gameplay, there is always a game \textbf{launch} stage lasting from tens of seconds to one minute, in the form of an opening animation streamed from the cloud server, to keep players waiting for the initialization of the game. In traditional gaming, such launch animations are installed and rendered on local gaming devices, while in cloud gaming, the animations that are made differently for each game title are live-streamed to players, leading to unique traffic characteristics across cloud game titles.

By analyzing our traffic traces labeled with their game titles, we observe unique packet group profiles in the downstream direction for their arrival times and payload sizes that carry launching scenes.
Fig.~\ref{fig:packet_genshin_win} visually shows the payload size and arrival time of the downstream packets in the RTP streaming flow for the first 60 seconds of a Genshin Impact gameplay session on a Windows PC with full HD graphics and 60fps streaming frame rate. From the scatter plot, we intuitively see that the packets fall into three groups, including \textbf{full} packets that have the same fixed (maximum) payload size and are constantly streamed to the local device; \textbf{steady} packets that hold similar payload sizes to their adjacent packets arriving within certain seconds; and \textbf{sparse} packets with high variations in payload sizes compared to their neighboring packets. 

The presence of the three packet groups holds similar profiles (\ie timing characteristics of each packet group and relative payload sizes across intervals of the same packet group) in all gameplay sessions of the same game titles, regardless of their client device types and streaming configurations. The packet statistics become unpredictable after the launching stage when players are involved in gameplay. As evidence, Fig.~\ref{fig:packet_genshin_android} and Fig.~\ref{fig:packet_genshin_win_hd} show the scatter plots for the same game title (\ie Genshin Impact) with different user settings. It is clear that the relative payload sizes and arrival time slots of the full, sparse, and steady packet groups in Fig.~\ref{fig:packet_genshin_android} remain almost identical compared to Fig.~\ref{fig:packet_genshin_win}. For Fig.~\ref{fig:packet_genshin_win_hd}, the arrival time slots of the three packet groups are identical compared to Fig.~\ref{fig:packet_genshin_win}, with tiny variations in relative payload sizes during three time slots (4.1 seconds in total) for the steady packet group.

Gameplay sessions of different game titles exhibit distinctive profiles for the three groups of packets. An example is provided in Fig.~\ref{fig:packet_fortnite_win} for a Fortnite gameplay session. The arrival density of full packets, the arrival time slots of sparse and steady packet groups, and their relative payload sizes are all different compared to game streaming sessions of the just-discussed Genshin Impact and other popular game titles in our dataset. 

\subsection{Flow Volumetric Profiles across Player Activity Stages} \label{sec:dissect_engagement_status}

After understanding the packet characteristics across game titles that indicate the overall QoE expectation of a game streaming session \cite{bhuyan_end_2022,sabet_delay_2020}, we now discuss the volumetric profiles of streaming flows that dynamically change on player activity stages in a cloud game. As a common practice in the gaming industry, graphics rendering and gaming computation are dynamically optimized at the software level for efficient resource consumption \cite{unity_reduce_rendering,xbox_graphics_effiency}. For cloud game streaming, such optimizations lead to reduced demands on QoE metrics (\eg streaming frame rate) and network resource consumption (\eg bandwidth) for delivering a good user experience in real time \cite{lyu_do_2024,marchal_analysis_2023}.

By analyzing the collected PCAP files of cloud gameplay with labeled player activities, we identify unique volumetric patterns in both upstream and downstream directions depending on the stages of player activity from idle, passive to active, with a visual example provided in Fig.~\ref{fig:volumetric_overwatch_win_hd}. The figure shows the upstream and downstream throughput of game streaming flows for an Overwatch gameplay session with background colors indicating the ground-truth player activity stages. 

\begin{figure}[t!]
	\mbox{
		\subfigure[Spectate-and-play games]{
			\includegraphics[width=0.22\textwidth]{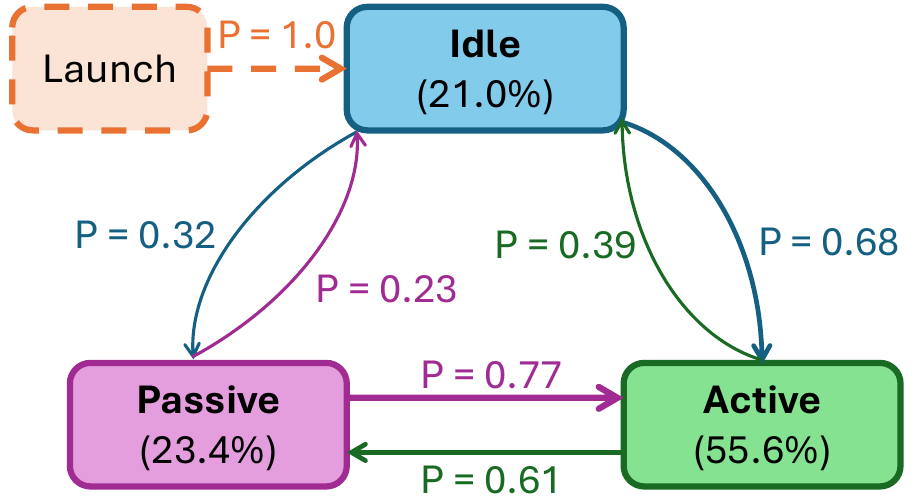}
			\label{fig:analyze_stage_transition_spectate}
		}
		\subfigure[Continuous-play games]{
			\includegraphics[width=0.22\textwidth]{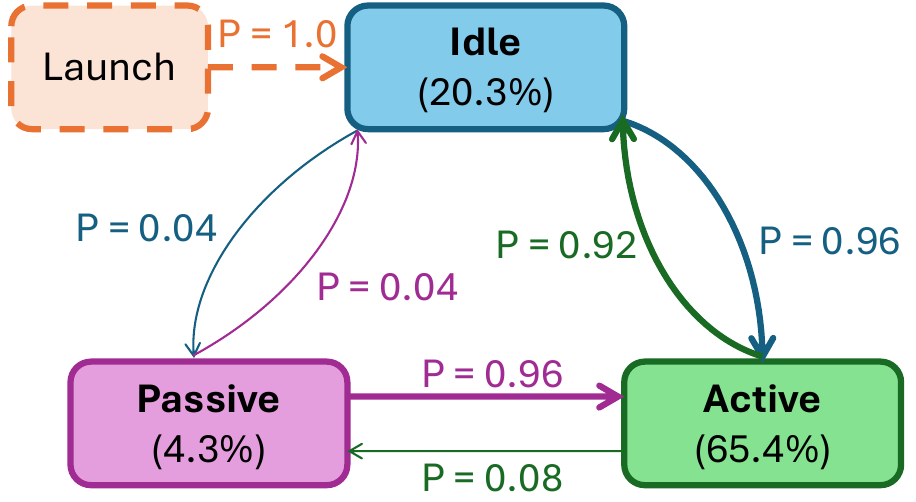}
			\label{fig:analyze_stage_transition_continuous}
		}
	}
   \vspace{-3mm}
	\caption{The average percentage of total playtime spent in the \coloridle{idle}, \colorpassive{passive}, and \coloractive{active} player activity stages with their transition probabilities per game streaming session for (a) spectate-and-play and (b) continuous-play cloud games in our labeled dataset.}
	\label{fig:analyze_stage_transition}
\end{figure}

For the \textbf{idle} player activity stage, we can see that the throughput in the downstream direction remains relatively low during the first idle stage when the player was ``browsing available matching options'' and drops to a low level in both directions when the player was ``back to the hub from the matching room'' at the end of this streaming session. 
In contrast, when the player activity stage becomes \textbf{active} for ``combating in progress'', the streaming throughput jumps to the highest level of the entire session, especially in the downstream direction.
During the \textbf{passive} stage when the player was passively watching the progress of gameplay such as ``being defeated by opponents in the match and watching the activity of teammates'', the downstream throughput remains at a similarly high level to the active stage but with reduced throughput in the upstream direction from the client to the cloud server. The observations in the relative (not absolute) changes of upstream and downstream volumetric profiles during a cloud gameplay session are consistent across all 13 popular game titles. Later in \S\ref{sec:detect}, we will leverage the bidirectional volumetric patterns as discussed to develop our method to reversely infer the player activity stages.

\begin{figure*}[t!]
    \includegraphics[width=\textwidth]{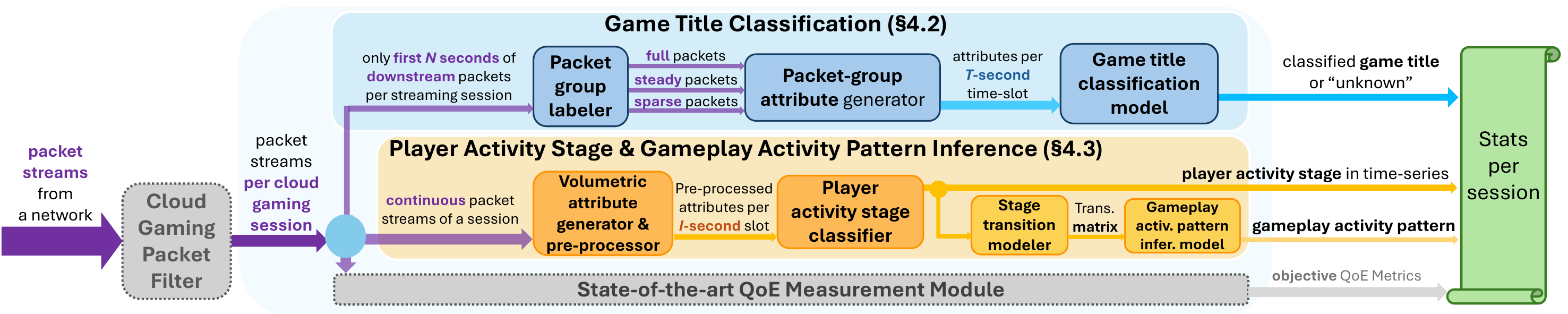}
    \vspace{-3mm}
	\caption{Our real-time network traffic analysis methodology to classify gameplay contexts of cloud game streaming sessions for effective user experience measurement.}
        \label{fig:pipeline}
\end{figure*}

As discussed in \S\ref{sec:cg_process}, due to the designed gameplay mechanism of each game title, the 13 popular games can be categorized by their gameplay activity patterns as either ``continuous-play'' or ``spectate-and-play''. We observe that the streaming sessions of the \textbf{continuous-play} game type, \eg Cyberpunk 2077 in Fig.~\ref{fig:volumetric_cyberpunk_mac_uhd}, often have a larger fraction of playtime in the active player activity stage compared to that of \textbf{spectate-and-play} games like Overwatch in Fig.~\ref{fig:volumetric_overwatch_win_uhd} and CS:GO in Fig.~\ref{fig:volumetric_csgo_win_uhd}. Also, the \textbf{transitions among player activity stages} hold similarities for gaming sessions of the same type of gameplay activity pattern. The two state transition diagrams in Fig.~\ref{fig:analyze_stage_transition} show the average fraction of playtime spent in the active, passive, and idle player activity stages per gameplay session, with their transition probabilities computed from all ground-truth sessions. For game streaming sessions of the ``spectate-and-play'' games, the active stage often takes about 40\% to 60\% of the total playtime and the passive stage accounts for the majority of the rest of the playtime. As for the ``continuous-play'' games, more than 95\% playtime is spent in the active or idle player activity stages and less than 5\% playtime is spent in the passive stage.

Our empirical findings in the differentiation of gameplay activity patterns show the ineffectiveness of existing cloud gaming user experience measurements relying solely on objective QoE metrics, which can mislabel ``spectate-and-play'' game streaming sessions as having poor user experience due to their low volumetric profiles during passive and idle stages. The findings also inspire us to leverage the transition behaviors of the player activity stages to classify the gameplay activity patterns.
\section{Classifying Cloud Game Contexts from Real-time Network Traffic}\label{sec:detect}
Driven by the insights into packet group characteristics and flow volumetric profiles across game titles and player activity stages in \S\ref{sec:dissect}, we develop our traffic analysis method (\S\ref{sec:method_overview}) to classify cloud game contexts for network operators in real time, including two classification processes that use statistical machine learning models to classify game titles (\S\ref{sec:classify_game_title}) and player activity stages (\S\ref{sec:classify_game_stage}). 
The classification performance of our method is extensively evaluated in our lab using ground-truth sessions (\S\ref{sec:evaluation}) before deployment in the wild.

\subsection{Methodology Overview}\label{sec:method_overview}

Our network traffic analysis method (overviewed in Fig.~\ref{fig:pipeline}) is designed to process real-time network traffic through an operational network to classify the gameplay contexts of cloud game streaming flows. 

Upon receiving real-time packet streams, the cloud gaming packet filter module selects only packets belonging to cloud game streaming flows for further processing and classification. We use adapted state-of-the-art cloud game streaming flow detection signatures \cite{lyu_network_2024,shirmarz_from_2024,graff_efficient_2023} that achieve 100\% accuracy from our lab validation in detecting RTP flows of four major cloud gaming platforms including NVIDIA's GeForce NOW, Microsoft's Xbox Cloud Gaming, Amazon Luna, and Sony's PS5 Cloud Streaming.

The packet streams of game streaming flows are then forwarded to three parallel processes. The bottom one, in the gray region of Fig.~\ref{fig:pipeline}, continuously measures objective QoE metrics, including game streaming frame rate, game streaming lag, and graphic resolution from flow QoS attributes using an established method by prior work \cite{lyu_network_2024}. The process highlighted in the light blue region detects game titles, while the one shown in the orange region continuously measures player activity stages and infers gameplay activity patterns. These two processes are novel in this work and will be discussed in detail. 

The first novel process analyzes the packet streams that belong to the first \textit{N} seconds of a game streaming flow to classify the respective game title being played, leveraging the packet group characteristics during the game launching stages. As will be discussed in \S\ref{sec:classify_game_title}, we select \textit{N = 5} for a balanced classification accuracy and responsiveness. The key technical components in this module including statistical attributes and classification models, will be elaborated soon in \S\ref{sec:classify_game_title}.

Driven by the insights of flow volumetric patterns across player activity stages from game launch to idle, active or passive, our second novel process (orange box in Fig.~\ref{fig:pipeline}) analyzes packet streams of both upstream and downstream directions in each game streaming flow to continuously label player activity stage. As will be discussed in \S\ref{sec:classify_game_stage}, upon receiving a sufficient number of past states and their transitions, this module will make a confident inference on the gameplay activity pattern of the session, which will be useful for estimating the QoE demands of a game streaming session when an inference result from the game title classification module does not fall into the known catalog. In our implementation, we operate both modules independently in parallel at their respective granularities, with the objective of cross-validating the classification results for each candidate game streaming session.

As the output of our method, the gameplay contexts will be correlated with state-of-the-art objective QoE metrics and/or network QoS attributes per streaming session, so that network operators can calibrate their measurement of user experience (which we refer to as effective QoE) to identify issues caused by network conditions rather than less demanding game titles and player activities. Later in \S\ref{sec:effective_qoe}, we will show an example of such context-based calibration deployed for our partnered network operator, which corrects mislabeled poor game streaming user experience due to low frame rate and throughput in less demanding scenarios.

\subsection{Classifying Game Titles}\label{sec:classify_game_title}

The game title classification process, as shown in the blue region of Fig.~\ref{fig:pipeline}, classifies the game title of a streaming session based on the first few seconds of its downstream packets delivering the game launch content. 
In \S\ref{sec:dissect_game_titles}, by analyzing the arrival time slots and payload sizes, we have empirically clustered three groups of \textbf{downstream} packets, namely full, steady, and sparse that arrive in the game launch stages. This drives our development of the game title classification process, which uses well-trained machine learning models on statistical attributes formulated per time slot from the three unique packet groups. Its key technical components are discussed below.

\subsubsection{Labeling downstream packet groups}\label{sec:label_packet_groups}
The downstream packets during the first \textbf{\textit{N}} seconds of each game streaming session will first be categorized into one of the three packet groups (\ie full, steady and sparse) by their relative payload sizes in the respective arrival time slot of \textbf{\textit{T}} seconds.
If \textit{T} is properly configured, the ``full'' packets that have the maximum payload size (\eg 1432 bytes) will consistently arrive in all time slots, as shown in Fig.~\ref{fig:packet_patterns}.
Additionally, in a \textit{T}-second time slot, packets that are not labeled as ``full'' will be labeled as either ``steady'' or ``sparse'' using a majority-voting method based on their payload sizes, with a tunable parameter \textbf{\textit{V}} that specifies the numerical range of the payload size variation for a packet compared to its adjacent packets. Specifically, a packet having a payload size within $\pm$\textit{V} difference than its neighbors within the time slot is identified as steady, otherwise sparse. This is driven by our findings in \S\ref{sec:dissect_game_titles} that each discrete time slot (\eg hundreds of milliseconds to tens of seconds) can contain steady packets densely located in one or several narrow bands of payload sizes; and/or sparse packets following a random distribution for their payload sizes. 
As will be discussed in \S\ref{sec:evaluate_title}, in our implementation, the packet group labeler achieves its best performance with the value of \textit{V} set to 10\% after evaluating a range of options from 1\% to 20\%.

\begin{figure}[!t]
    \centering
    \includegraphics[width=\columnwidth]{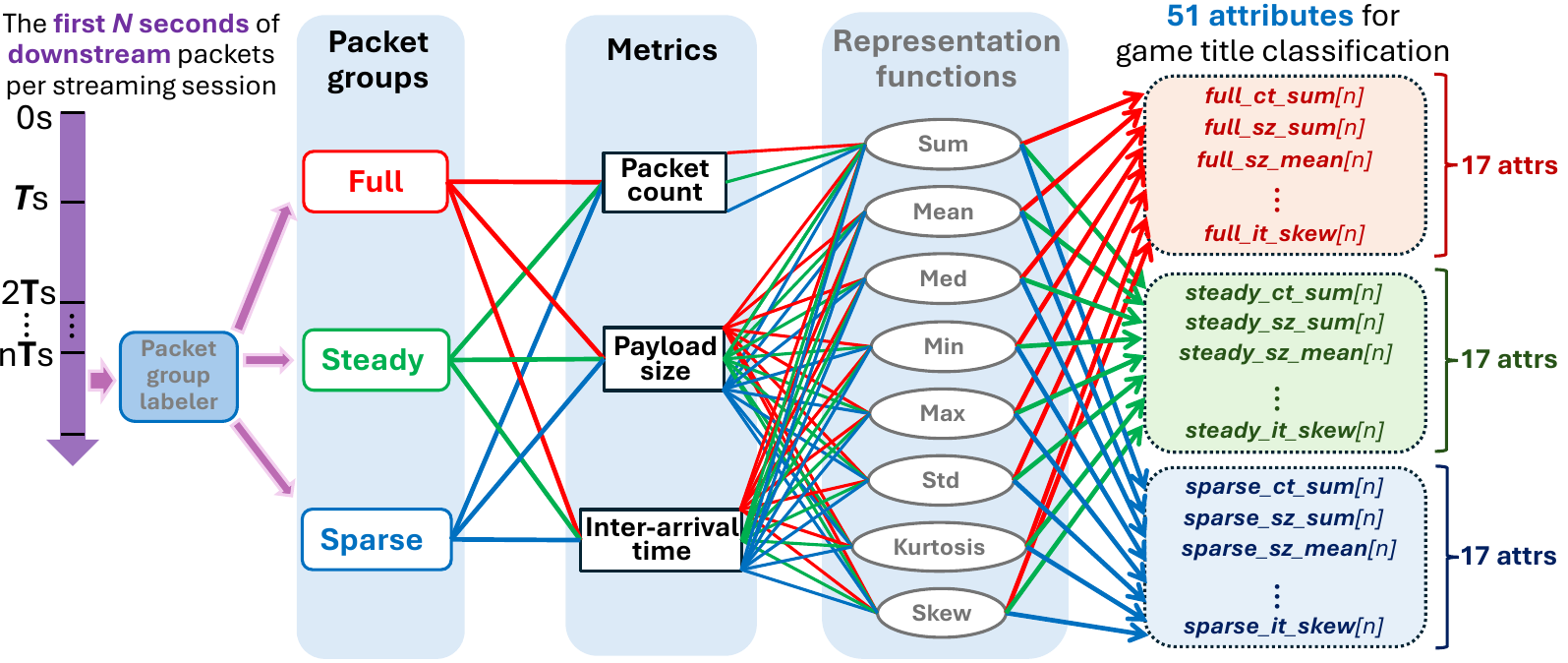}
    \vspace{-3mm}
    \caption{Formulating attributes from the full, steady and sparse packet groups per time slot.}
    \label{fig:packet_group_attributes}
\end{figure}

\subsubsection{Attributes of packet groups per time slot}\label{sec:title_attr}
Packets labeled full, steady, or sparse are then used to generate statistical attributes per packet group for the respective \textit{T}-second time slot.
Fig.~\ref{fig:packet_group_attributes} visually depicts the formulation of our per-time-slot attributes describing statistical representations of packet count, payload size, and inter-arrival time for the packets within the ``full'', ``steady'', or ``sparse'' group.
For example, the attribute annotated as \textit{full\_ct\_sum[0]} shows the total count of packets in the ``full'' group during the first \textit{T}-second time slot of this game launching stage. As will be discussed in \S\ref{sec:evaluation}, in our implementation, we process the first 5 seconds of packets in a 1-second time slot to achieve good classification performance.

\subsubsection{Classification model}\label{sec:title_model}
For each game streaming session, the attributes generated from all \textit{T}-second time slots for the first \textit{N} seconds will be processed in a batched manner by our classification model. The model will predict a (popular) game title that exists in our pre-trained dataset or the ``unknown'' type if a confident result cannot be made. In \S\ref{sec:evaluation}, we discuss our evaluation of the classifier trained with common machine learning algorithms, with the best-performing random-forest classifier selected in our implementation.

We acknowledge that flow classification for network operators is often expected to be as early as possible for subsequent network operational tasks \cite{babaria_fastflow_2025,piet_ggfast_2023}. In this work, our game title classification takes only a few seconds (\ie 5 seconds in our implementation) from the game launch stage, allowing the network operator to seamlessly start subsequent measurement tasks and/or prioritize game streaming flows of high-demanding games prior to the start of actual gameplay.

\subsection{Classifying Player Activity Stages and Inferring Gameplay Activity Patterns} \label{sec:classify_game_stage}
The second novel process in our method, as outlined in the yellow region of Fig.~\ref{fig:pipeline}, continuously classifies the player activity stages of a game streaming session using its bidirectional volumetric attributes and infers the activity pattern of this gameplay from the transition pattern of the classified player activity stages.

\subsubsection{Classifying player activity stages}\label{sec:classify_engagement_status}
This process first classifies and tracks the player activity stages as active, passive, or idle over time by processing the bidirectional packet streams of the cloud game streaming session. Standard volumetric attributes for a game streaming session, including packet rate and throughput in both upstream and downstream directions, are computed per time slot of \textit{\textbf{I}} second, \eg one second in our implementation after balancing classification responsiveness and accuracy.

As discussed in \S\ref{sec:dissect_engagement_status}, each game streaming session can exhibit different absolute value ranges of their volumetric attributes across the three player activity stages, while their relative volumetric levels remain consistent regardless of the actual game titles and streaming configurations. Therefore, in each \textit{\textbf{I}}-second classification time slot, the four volumetric attributes are converted to the relative fraction compared to their peak value (above a threshold dynamically decided during the game launch) observed from previous slots.

To eliminate the effect of noisy attribute values caused by unexpected short behavior for a certain type of player activity stage, such as accidental mouse movement (\ie high upstream throughput) when spectating teammates (\ie passive stage), which will lead to a classification result as active for this slot rather than the correct passive stage, the volumetric attributes are processed by exponential moving average (EMA) to consider the attribute values from the past slots as shown in Equation~\ref{eq:EMA}. The weight of the current stage $\alpha$ is selected as 0.4 from the range of (0,1) for the best accuracy, as will be evaluated soon in \S\ref{sec:evaluation}.

\begin{equation}
\text{attr}_t = \alpha \cdot \text{attr}_t + (1 - \alpha) \cdot \text{attr}_{t-1}
\label{eq:EMA}
\end{equation}

A well-trained machine learning classifier developed using the Random Forest algorithm then consumes preprocessed volumetric attributes in their EMA-smoothed relative values to classify player activity stages as idle, passive, or active for each slot accordingly.

\subsubsection{Inferring gameplay activity pattern} \label{sec:classify_activity_pattern}
As discussed in \S\ref{sec:cg_process}, depending on the game titles, cloud game sessions exhibit two distinctive patterns for their gameplay activities, namely ``continuous-play'' and ``spectate-and-play''. 
If the fine-grained game title cannot be confidently classified for a game streaming session, our method can provide network operators with visibility into the gameplay activity pattern, revealing coarse-grained profiles and network demands for the respective game streaming sessions.

As quantitatively shown in Fig.~\ref{fig:analyze_stage_transition}, the two types of gameplay activity patterns hold different proportions of player activity stages and transition probabilities between these stages.
Therefore, our inference model captures the stochastic transition behaviors between player activity stages in a gameplay for classification.

As shown in the process in Fig.~\ref{fig:pipeline}, the continuously classified player activity stages are also received by the stage transition modeler. In our implementation, for each game streaming session, we use a 3x3 matrix with each cell representing the occurrence of transition per slot from one stage to another or retaining itself. The inference model (\eg using random forest algorithm) makes a prediction based on the nine values of the matrix normalized to their probabilities across time slots within the monitored duration when its confidence level is above a threshold (\eg 75\%) tuned for balanced prediction responsiveness and accuracy. 

\subsection{Parameter and Model Evaluation}\label{sec:evaluation}
After having a comprehensive understanding of our network traffic analysis method for cloud gameplay context classification as depicted in Fig.~\ref{fig:pipeline}, we now evaluate the options for inference models and key parameters in the pipeline, as well as its overall classification performance for game titles, player activity stages and gameplay activity patterns. The lab evaluation uses our ground-truth dataset (described in \S\ref{sec:dataset}). Similar to prior works \cite{babaria_fastflow_2025,wang_data_2024,jiang_netdiffusion_2024} in traffic classification, we augment our dataset for larger sample sizes using variation-based statistical techniques, \ie by synthesizing packet data with randomly varied sizes and arrival times based on the original ground-truth data, especially for classes with fewer samples. We now present evaluation results for key design choices in developing the game title and player activity stage classification processes.

\subsubsection{Evaluating game title classification}\label{sec:evaluate_title}
As discussed in \S\ref{sec:classify_game_title}, the game title classification uses three tunable parameters \textbf{\textit{N}}, \textbf{\textit{T}} and \textbf{\textit{V}}, along with a machine learning classifier, which were selected for our implemented system after lab evaluations. 

\begin{figure}[t!]
    \centering
    \subfigure[0.1-second time slot.]{
        \includegraphics[width=0.475\columnwidth]{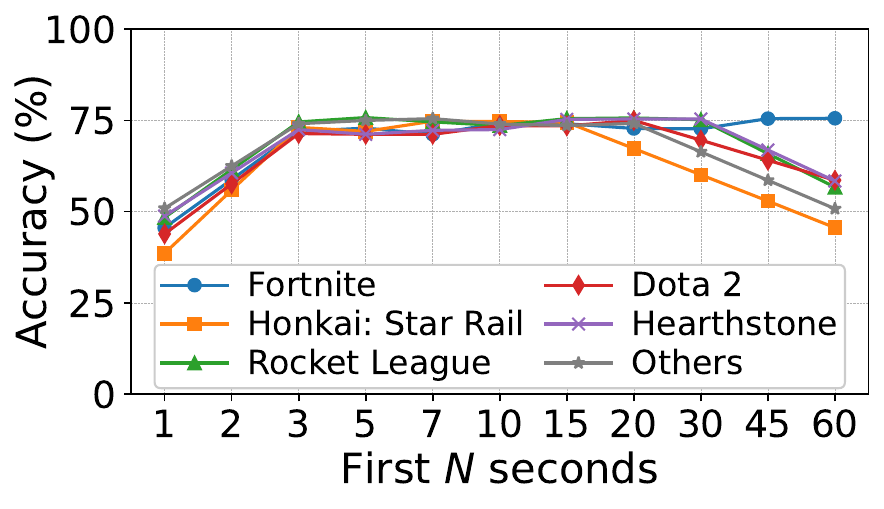}
        \label{fig:game_title_tuning_time_slot_100ms}
    }
    \subfigure[0.5-second time slot.]{
        \includegraphics[width=0.475\columnwidth]{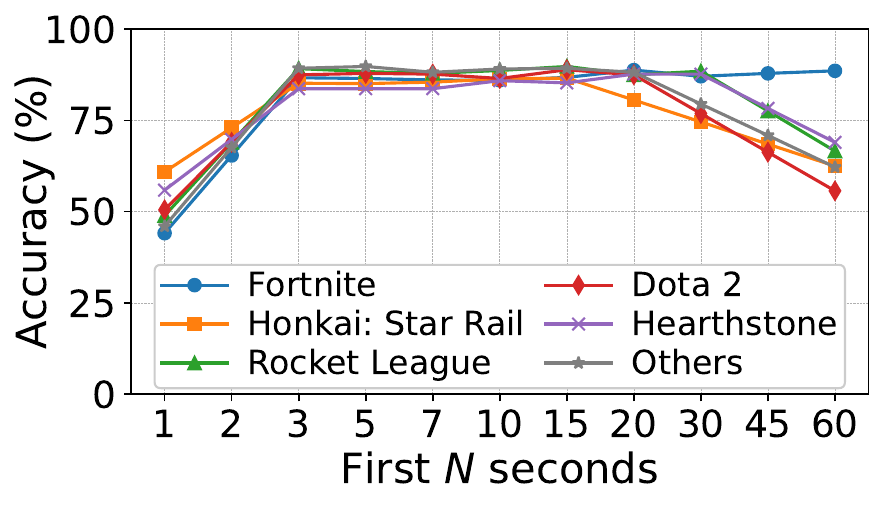}
        \label{fig:game_title_tuning_time_slot_500ms}
    }
    \subfigure[1-second time slot.]{
        \includegraphics[width=0.475\columnwidth]{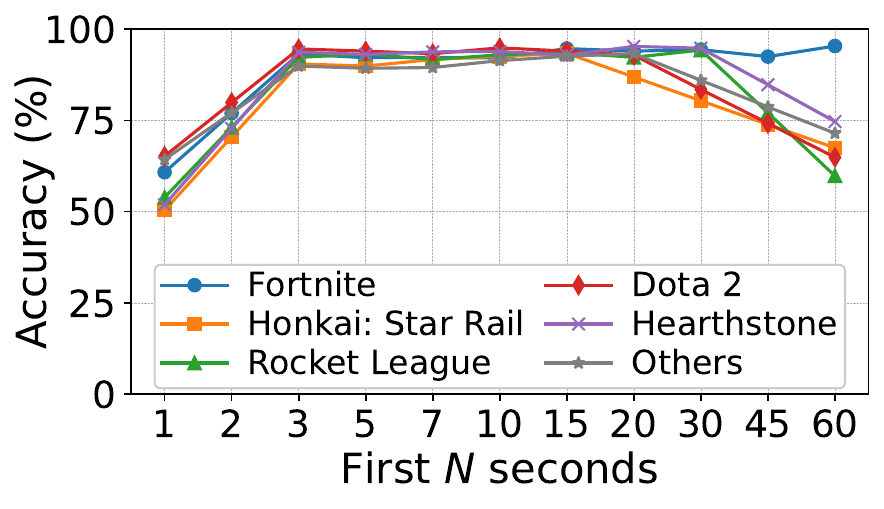}
        \label{fig:game_title_tuning_time_slot_1000ms}
    }
    \subfigure[2-second time slot.]{
        \includegraphics[width=0.475\columnwidth]{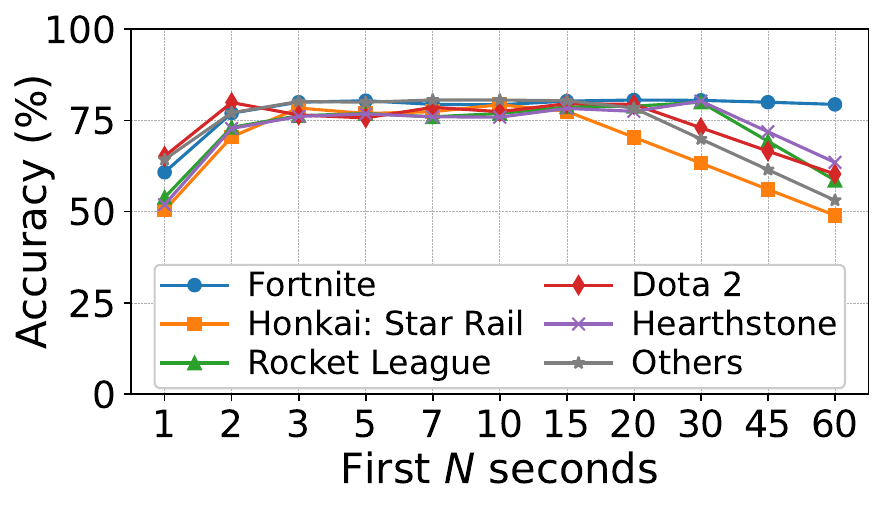}
        \label{fig:game_title_tuning_time_slot_2000ms}
    }
    \vspace{-3mm}
    \caption{Accuracy of game title classification using attributes computed from packets from the first 1 to 60 seconds of game launch with time slot sizes as (a) 0.1 seconds, (b) 0.5 seconds, (c) 1 second, and (d) 2 seconds.}
    \label{fig:game_title_tuning_time_slot}
\end{figure}

\textbf{Time window \textit{N} and time slot \textit{T}:}
Our attributes are calculated from the packet statistics per time slot \textbf{\textit{T}} within the first \textbf{\textit{N}} seconds of the game launch stage. A sufficiently large time window \textit{N} allows more identical packet statistics to be captured by the attributes. Similarly, a properly selected time slot \textit{T} helps the attributes better profile the statistical features while minimizing the impact of mild delays and packet loss that can be diluted in a sufficiently large time slot. Therefore, we tune the best-performing classification models (processes to be discussed later) for different options of \textit{T} and \textit{N}, with their accuracy for five representative game titles provided in Fig.~\ref{fig:game_title_tuning_time_slot}.
It is clear that the classification accuracy follows an increasing trend for both \textit{T} and \textit{N} until a certain value, \ie 3 seconds for \textit{T} and 1 second for \textit{N} with over 95\% accuracy on average, which are the suitable options to be used in our system implementation. Notably, we use the downstream packets from the first five (instead of three) seconds of a game streaming session to accommodate possible packet arrival delays in the game launch stage.
It is worth noting that the implemented values of the tunable parameters \textit{N} and \textit{T} are selected based on our lab dataset without additionally introduced latency or packet loss. We acknowledge that different values can be selected for extended datasets with varying network conditions, which is beyond our current scope.

\textbf{Payload size variation \textit{V} for packet group labeling: }
While full packets can be simply labeled by those having maximum payload size, to precisely distinguish steady and sparse packets for game title classification, we tune the parameter \textit{V} used in our majority-voting algorithm as discussed in \S\ref{sec:label_packet_groups}, which represents the range of payload size variation allowed among adjacent steady packets in a time slot.
Since sessions of a certain game title show consistent packet patterns, we randomly pick one representative session from each of the thirteen games and visually inspect the labeled packet groups using \textbf{\textit{V}} between 1\% and 20\%.
From our evaluation results, we conclude that a high range of allowed variation (\ie 15\% and 20\%) leads to sparse packets being mistakenly labeled as steady packets, while low variation (\ie 1\% and 5\%) mislabels certain steady packets with only slight discrepancy as sparse packets.
The observation is consistent across all thirteen games, therefore, we use \textit{V} of 10\% in our implementation, which yields the best labeling results.

\begin{table}[t!]
    \centering
        \caption{Game title classification accuracy of the best-performing classifier using our specialized packet-group-based attributes vs standard flow volumetric attributes.}
        \vspace{-3mm}
    \resizebox{\columnwidth}{!}{
    \begin{tabular}{|l|l|l|}
    \hline
    \rowcolor[HTML]{C0C0C0} 
    \textbf{Game title}                               & \textbf{Accur. (pkt. group)} & \textbf{Accur. (flow vol.)} \\ \hline
    {\color[HTML]{3531FF} \textbf{Baldur's Gate 3}}   & 96.9\%                                & 82.1\%                                    \\ \hline
    {\color[HTML]{3531FF} \textbf{Cyberpunk 2077}}    & 94.9\%                                & 81.5\%                                    \\ \hline
    {\color[HTML]{3531FF} \textbf{Genshin Impact}}    & 98.0\%                                & 89.8\%                                    \\ \hline
    {\color[HTML]{3531FF} \textbf{Honkai: Star Rail}} & 94.4\%                                & 85.9\%                                    \\ \hline
    {\color[HTML]{6200C9} \textbf{Call of Duty}}      & 95.2\%                                & 80.5\%                                    \\ \hline
    {\color[HTML]{6200C9} \textbf{CSGO}}          & 97.5\%                                & 81.8\%                                    \\ \hline
    {\color[HTML]{6200C9} \textbf{Destiny 2}}         & 94.1\%                                & 86.0\%                                    \\ \hline
    {\color[HTML]{6200C9} \textbf{DOTA 2}}            & 92.7\%                                & 84.9\%                                    \\ \hline
    {\color[HTML]{6200C9} \textbf{Fortnite}}          & 97.3\%                                & 91.5\%                                    \\ \hline
    {\color[HTML]{6200C9} \textbf{Hearthstone}}       & 94.8\%                                & 88.0\%                                    \\ \hline
    {\color[HTML]{6200C9} \textbf{Overwatch 2}}       & 95.6\%                                & 85.4\%                                    \\ \hline
    {\color[HTML]{6200C9} \textbf{R6: Siege}}         & 94.3\%                                & 83.7\%                                    \\ \hline
    {\color[HTML]{6200C9} \textbf{Rocket League}}     & 92.8\%                                & 83.5\%                                    \\ \hline
    \end{tabular}
    }
    \label{tab:evaluate_title}
\end{table}

\begin{figure}[t!]
	\includegraphics[width=\columnwidth]{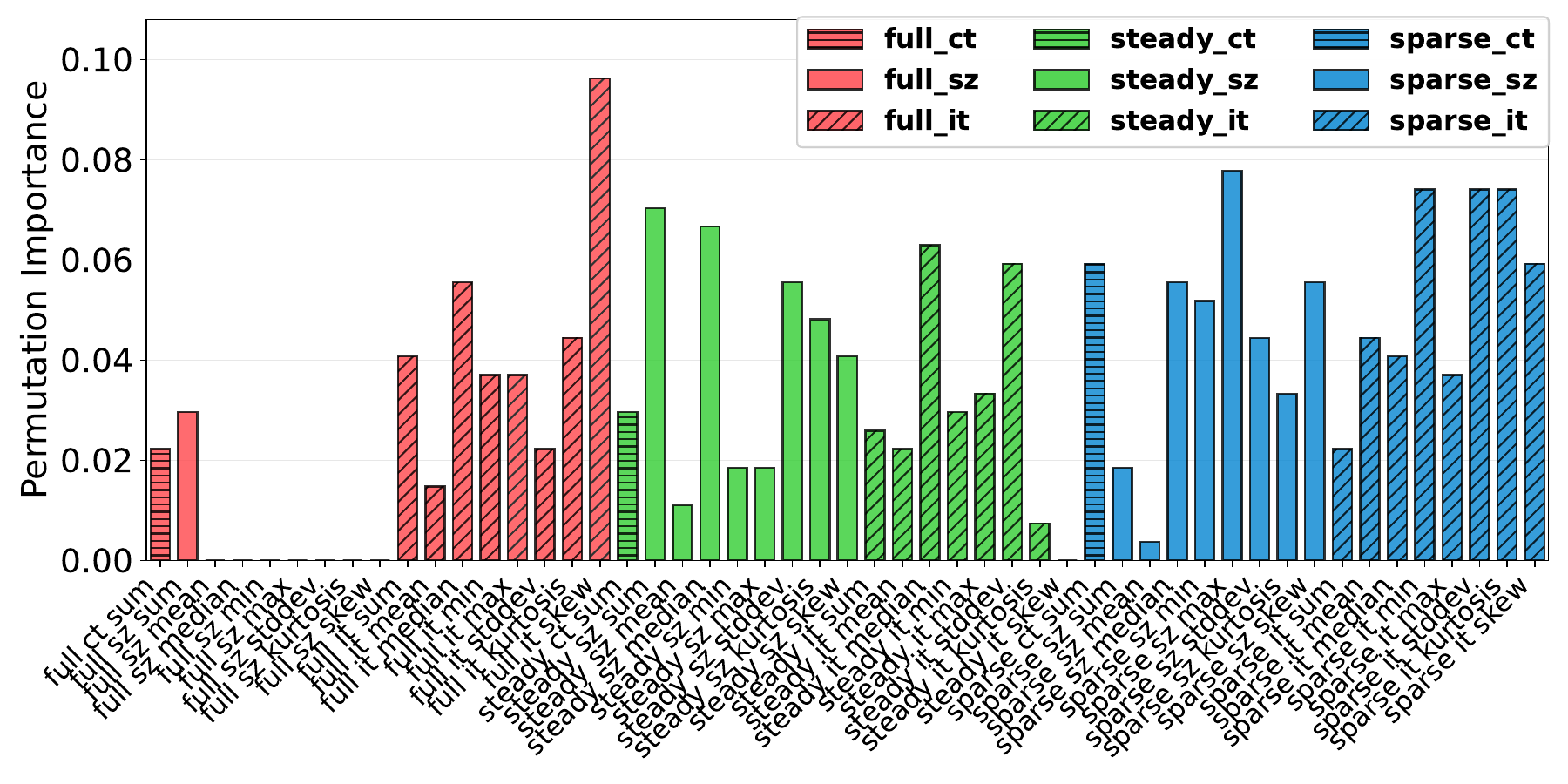}
	\vspace{-8mm}
\caption{Importance of the 51 attributes in classifying game titles by the best-performing Random Forest model. Each of them is color-coded by its packet group (full, steady, or sparse) and pattern-coded by the metric (count, size, or inter-arrival time).}
\label{fig:game_title_feature_importance}
\end{figure}

\textbf{Machine learning classifier:}
To select a suitable machine learning model as our game title classifier, we fine-tune and evaluate three commonly used machine learning algorithms, namely margin-based Support Vector Machine (SVM), distance-based K-Nearest Neighbors (KNN), and Random Forest (RF).
After fine-tuning the hyperparameters of each algorithm -- such as the regularization and kernel type for SVM, the number of neighbors and distance metric for KNN, and the number of trees and maximum tree depth for random forest -- the classifier that achieves the best overall accuracy (\ie over 95\%) in classifying all game titles is selected, with its accuracy for each game title provided in the second column of Table~\ref{tab:evaluate_title}. 
The details of our hyperparameter tuning and model evaluation are provided in Appendix \S\ref{sec:additional_title_eval}.

We evaluate the importance of the attributes consumed by the selected random forest classifier using the permutation importance metric \cite{breiman_random_2001}, which measures the drop in the accuracy of the model when the values of an evaluated attribute are shuffled randomly. As visually shown in Fig.~\ref{fig:game_title_feature_importance}, 43 attributes exhibit certain predictive powers in classifying game titles, while the other eight attributes, including seven for the full packet group and one for the steady group, have little significance with their measured importance being 0. Unsurprisingly, the eight attributes describe less varying statistics among game titles, such as the mean packet sizes of the full packet group, which can be excluded in the classification pipeline to optimize the processing cost \cite{wan_cato_2025}.

For comparison, we also trained classifiers that take the two standard flow volumetric attributes (\ie packet rate and throughput) per time interval rather than our well-articulated attributes from the three packet groups. Not surprisingly, the accuracy drops significantly (\eg more than 10\%) for each game title, as shown in the rightmost column of Table~\ref{tab:evaluate_title}. 
We have observed that most misclassified game steaming sessions have label confidence less than 40\%. Therefore, in our implementation, sessions with low-confidence results are labeled as ``unknown'' game title, suggesting that network operators should include more game titles in their databases if they wish, or refer to the coarse-grained gameplay activity patterns for those streaming sessions.

\begin{figure}[t!]
\vspace{-2mm}
    \centering
    \subfigure[0.1-second time slot.]{
        \includegraphics[width=0.47\columnwidth]{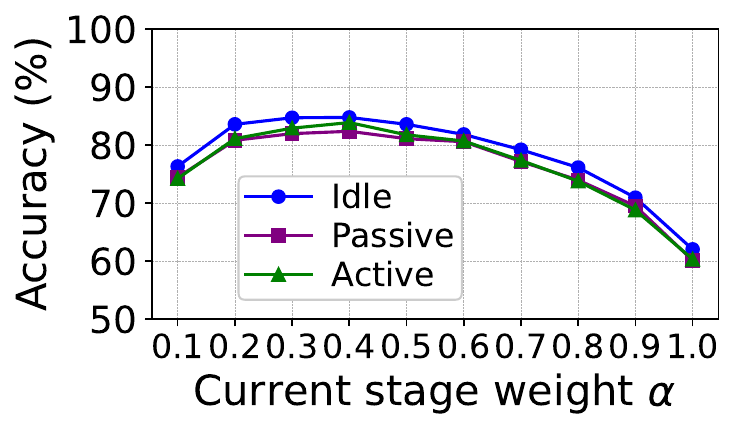}
        \label{fig:alpha_accuracy_time_slot_100ms}
    }
    \subfigure[0.5-second time slot.]{
        \includegraphics[width=0.47\columnwidth]{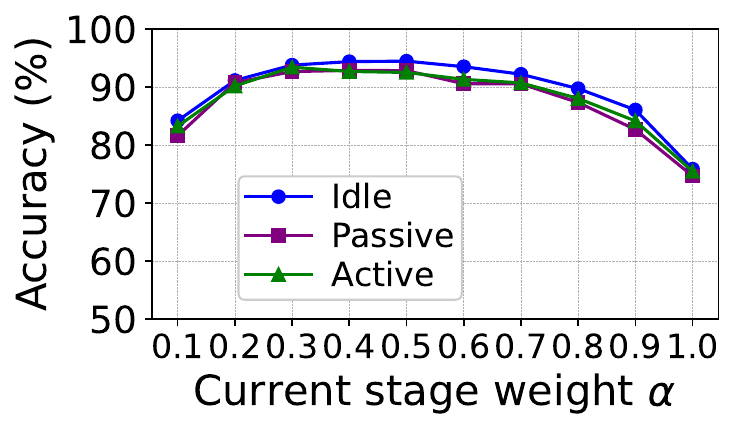}
        \label{fig:alpha_accuracy_time_slot_500ms}
    }
    \subfigure[1-second time slot.]{
        \includegraphics[width=0.47\columnwidth]{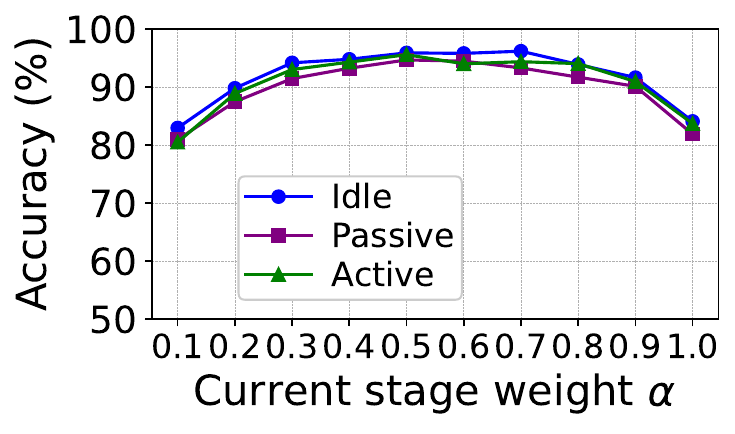}
        \label{fig:alpha_accuracy_time_slot_1000ms}
    }
    \subfigure[2-second time slot.]{
        \includegraphics[width=0.47\columnwidth]{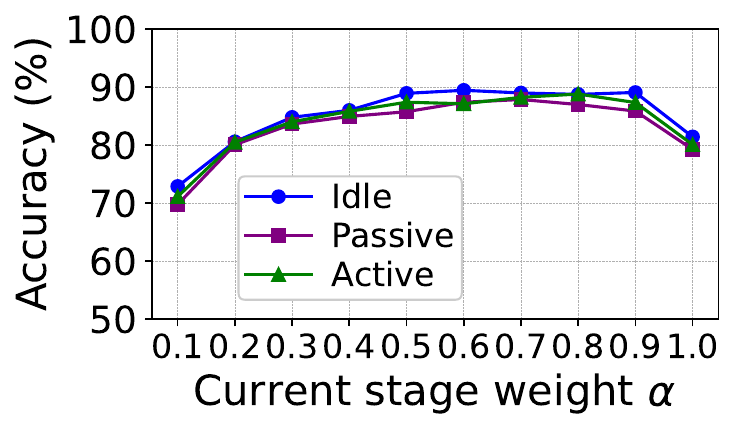}
        \label{fig:alpha_accuracy_time_slot_2000ms}
    }
    \vspace{-3mm}
    \caption{Accuracy of player activity stage classification using current stage weights between 0.1 and 1 with time slot sizes of (a) 0.1 seconds, (b) 0.5 seconds, (c) 1 second, and (d) 2 seconds.}
    \label{fig:alpha_accuracy_time_slot}
    \vspace{-5mm}
\end{figure}

\subsubsection{Evaluating player activity stage classification and gameplay activity pattern inference}\label{sec:evaluate_stage}
Our second process (\S\ref{sec:classify_game_stage}) continuously classifies the player activity stage and infers the gameplay activity pattern of a streaming session. We now discuss the selection of key options, including the time slot \textit{\textbf{I}} for player activity stage classification, the weight of the current activity stage \textit{\textbf{$\alpha$}}, the confidence threshold for outputting gameplay activity pattern inference results, and the machine learning classification models.

\textbf{Time slot \textit{I} and current stage weight $\alpha$:}
In the process that continuously classifies the player activity stage in a cloud game session, as discussed in \S\ref{sec:classify_engagement_status}, the intermediate labels are produced per \textit{I}-second time slot and smoothed by an exponential moving average (EMA) function with the weight of the current stage as $\alpha$. Therefore, we evaluate the accuracy produced by our best-performing classifiers that use different options of the two parameters, with representative results shown in Fig.~\ref{fig:alpha_accuracy_time_slot}. We can see in Fig.~\ref{fig:alpha_accuracy_time_slot_1000ms} that classification accuracy reaches its best performance with a 1-second time slot, as a small time slot can be overly granular to capture identical volumetric characteristics of the respective player activity stage, while a large time slot can be mixed with multiple stages. Regarding the current stage weight $\alpha$, a value between 0.5 and 0.6 has the best overall performance. In our implementation, we use a 1-second time slot with an $\alpha$ value of 0.5 to achieve classification robustness to short noises in streaming volumetric profiles.

\textbf{Confidence threshold for gameplay activity pattern inference:}
In our second process, we continuously infer the gameplay activity pattern based on the stage transition/retention rates upon each newly classified stage at a time slot, until the confidence level of the inferred pattern exceeds a certain threshold.
To balance accuracy (by using more data from more time slots) and responsiveness (\ie shorter inference time), we evaluate the per-session average accuracy and time needed to generate a confident inference result above confidence thresholds from 0\% up to 95\% for continuous-play and spectate-and-play games, respectively, using the optimal settings for other parameters.
By using very low confidence thresholds (\eg 0\% to 40\%), the model tends to generate results very early (\eg 1 to 30 seconds) but they are highly inaccurate (\ie less than 50\% accuracy).
On the other hand, overly high thresholds (\eg 95\%) may lead to inference results not being generated until close to the end of sessions, which are not useful for real-time monitoring and performance boost.
Therefore, we choose 75\% as the confidence threshold since it produces results with around 90\% accuracy for both types of games while only needing five minutes on average, at which time the player has just begun the gameplay.

\textbf{Machine learning classifiers:} In this process, we have developed two machine learning models to classify the player activity stage in a cloud game session using bidirectional volumetric attributes of the streaming session and to infer the overall gameplay activity pattern. We follow a similar process as described in \S\ref{sec:evaluate_title}, evaluating commonly used machine learning algorithms and their hyperparameters to select the best-performing models. 
To classify player activity stages using volumetric attributes, both throughput and packet rate demonstrate similar importance. The downstream attributes contribute more to the classification of the idle stage, while the upstream attributes are more important for the active and passive stages.
To classify gameplay activity patterns using the stage transition behaviors, the attributes that capture transitions from active to idle stages and from idle to passive stages are the most important ones. The details of our model and attribute evaluation are provided in Appendix \S\ref{sec:additional_pattern_eval}.
The accuracies of the best-performing models for both classification tasks are provided in Table~\ref{tab:activity_stage}. We observe that decent accuracy (\eg mostly over 95\%) is achieved for all classification labels in both player activity stages (\ie active, passive and idle) and gameplay activity pattern types (\ie continuous-play and spectate-and-play), meeting the requirement for the field deployment in our partnered ISP network as discussed next.

\section{Measuring Cloud Gaming Contexts at a Large Scale}  \label{sec:insights}

We have implemented and deployed our cloud gaming context measurement method in our partner ISP that hosts NVIDIA's GeForce NOW cloud gaming servers for our geography, providing valuable insights to precisely identify and troubleshoot user experience degradations that are indeed caused by network factors rather than mistaking low-throughput sessions caused by game titles and gameplay activity patterns as having poor experience. 
The real-time inference of game contexts is also meant to help our partner ISP in assessing the efficacy of their prioritization techniques in assuring cloud gaming experience over their 5G mobile broadband network.

\newcolumntype{?}[1]{!{\vrule width #1}}

\begin{table}[t!]
    \centering
        \caption{Player activity stage classification (by time slot) and gameplay activity pattern inference (by session) accuracy for continuous-play and spectate-and-play games.}
       \vspace{-3mm}
    \resizebox{\columnwidth}{!}{
    \begin{tabular}{|l|l?{0.5mm}l|l|}
    \hline
    \rowcolor[HTML]{C0C0C0} 
    \textbf{Gameplay actv. pattern}                                         & \textbf{Accur.} & \textbf{Player actv. stage} & \textbf{Accur.} \\ \hline
    {\color[HTML]{3531FF} }                                             &                                      & \coloractive{\textbf{Active}}                    & 94.1\%                             \\ \cline{3-4} 
    {\color[HTML]{3531FF} }                                             &                                      & \colorpassive{\textbf{Passive}}                   & 92.5\%                             \\ \cline{3-4} 
    \multirow{-3}{*}{{\color[HTML]{3531FF} \textbf{Continuous-play}}}   & \multirow{-3}{*}{95.7\%}             & \coloridle{\textbf{Idle}}                      & 97.6\%                             \\ \hline
    {\color[HTML]{6200C9} }                                             &                                      & \coloractive{\textbf{Active}}                    & 96.8\%                             \\ \cline{3-4} 
    {\color[HTML]{6200C9} }                                             &                                      & \colorpassive{\textbf{Passive}}                   & 95.9\%                             \\ \cline{3-4} 
    \multirow{-3}{*}{{\color[HTML]{6200C9} \textbf{Spectate-and-play}}} & \multirow{-3}{*}{97.2\%}             & \coloridle{\textbf{Idle}}                      & 98.4\%                             \\ \hline
    \end{tabular}
    }
    \label{tab:activity_stage}
\end{table}

In this section, we discuss representative insights from a three-month (1 December 2024 to 1 March 2025) measurement of cloud game contexts. The results are selected to comply with commercial confidentiality restrictions and our university ethical approval conditions as provided in Appendix~\S\ref{sec:appendix_ethics}.
One month before our field deployment, we evaluated our game title classification method in the deployment network by validating our classified game titles with the offline cloud game server logs that were generated after the conclusion of each game session and are not available to ISPs in daily operations. The evaluation results showed that our method can effectively classify the game titles for the thirteen popular games with an overall accuracy above 95\%, consistent with our lab evaluation results. As for the player activity stages, there is no record from cloud game server logs for us to validate outside the lab environment.

We now demonstrate how our measurement of cloud game contexts can help network operators understand player activity profiles across game titles (\S\ref{sec:usage_patterns}) which can lead to different expectations on game streaming quality; benchmark bandwidth demands that are unique to game titles and gameplay activity patterns (\S\ref{sec:network_demands}); and measure effective cloud gameplay experience by corroborating gameplay contexts with objective QoE and network QoS metrics (\S\ref{sec:effective_qoe}).

\begin{figure}[t!]
	\centering
	\subfigure[Average durations of player activity stages across 13 popular game titles.]{
		\includegraphics[width=\columnwidth]{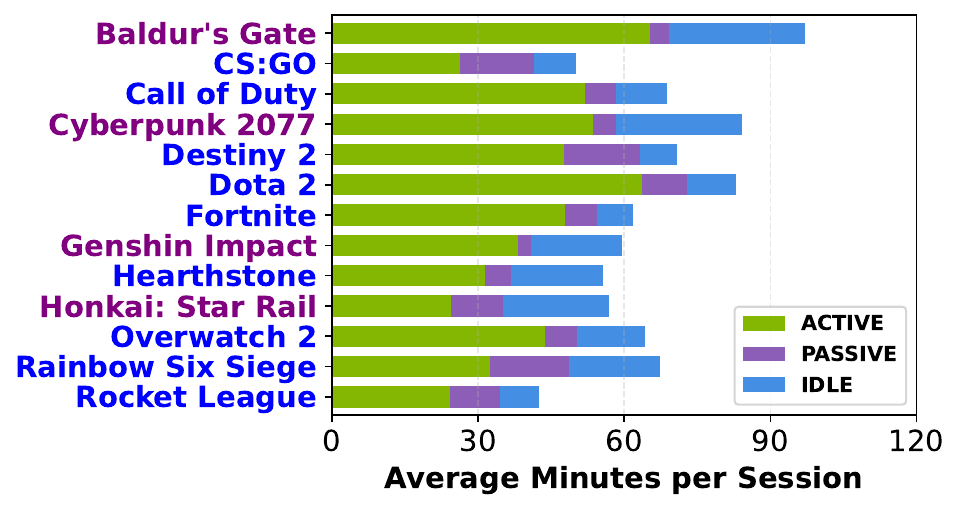}
		\label{fig:average_durations_barh}
	}
	\subfigure[Average durations of player activity stages across 2 activity patterns.]{
		\includegraphics[width=\columnwidth]{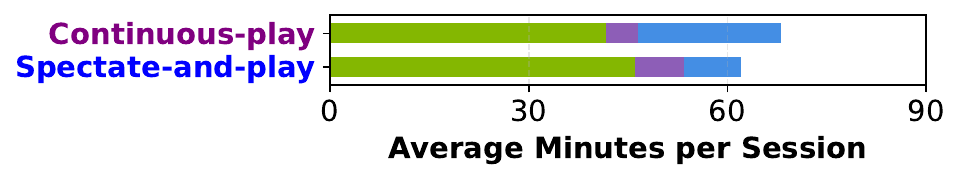}
		\label{fig:average_durations_barh_by_type}
	}
   \vspace{-3mm}
	\caption{Average number of minutes spent in \coloractive{active}, \colorpassive{passive}, and \coloridle{idle} player activity stages per cloud gaming session for (a) 13 popular game titles and (b) 2 gameplay activity pattern types.}
	\label{fig:average_durations}
\end{figure}

\subsection{Player Activity Stages across Cloud Game Contexts}\label{sec:usage_patterns}
Our measurement of cloud gaming contexts provides network operators with player activity profiles of streaming sessions per cloud game title and genre, which serve as references for network operators to dynamically provision network resources, \eg allocate 5G eMBB slices with prioritized QoS profiles by our partnered ISP with an expected session duration and slice capacity, upon detecting a newly commenced game streaming session. 

We report the average duration of each game streaming session and each player activity stage (\ie active, passive, and idle) across the 13 popular game titles in Fig.~\ref{fig:average_durations_barh}. The results are aggregated for the entire three-month deployment period. We can clearly observe significant differences in both streaming durations and player activity stages.

For the session duration, nine games have long average durations near or above 1 hour. Those game titles with long session durations are either role-playing games that involve extensive dialogue-based content like Baldur's Gate (95 minutes) and Cyberpunk 2077 (82 minutes), or MOBA, Sports and Shooter games like Dota 2 and Rainbow Six Siege that have either long matchmaking or match durations. 
Rocket League and CS:GO are with short per-match duration and have the shortest average game streaming durations among the popular games.
Depending on the gameplay content, each game title has its unique distribution of player activity stages, such as a relatively larger fraction (\eg 25 -- 55\%) of duration in idle and passive stages for Baldur's Gate, Cyberpunk 2077, Honkai: Star Rail and Hearthstone due to their extensive static content or unskippable dialogues. On the contrary, Fortnite and Dota 2 have most of their gameplay duration in the active stage. 

For game sessions that do not belong to the thirteen popular titles and thus are coarsely categorized as either continuous-play or spectate-and-play games, from Fig.~\ref{fig:average_durations_barh_by_type}, we can also observe different expectations on the streaming session durations and the composition of player activity stages. Continuous-play games, which are often role-playing, have a longer session duration with a significant fraction of time spent in idle (26\%) stages. Spectate-and-play games are often match-based shooter, card, strategy or sports games without lengthy in-game static scenes and dialogues, therefore, have a larger fraction of time spent in the active stage.

\subsection{Network Bandwidth Demands across Cloud Game Contexts}   \label{sec:network_demands}
Among the network resource provisioning tasks, ensuring sufficient bandwidth for multimedia streaming sessions to support good quality-of-experience (QoE) metrics is often very challenging for network operators who operate with constrained bandwidth resources for massive subscribers, especially in high-density regions (\eg a 5G base station serving a metropolitan area) during peak hours.
Cloud game streaming sessions with different contexts exhibit various network bandwidth demands, determined not only by their streaming settings but also by the gameplay activity patterns that are inherently different across game titles.

\begin{figure}[t!]
    \centering
    \subfigure[Average throughput per session across 13 popular game titles.]{\includegraphics[width=\columnwidth]{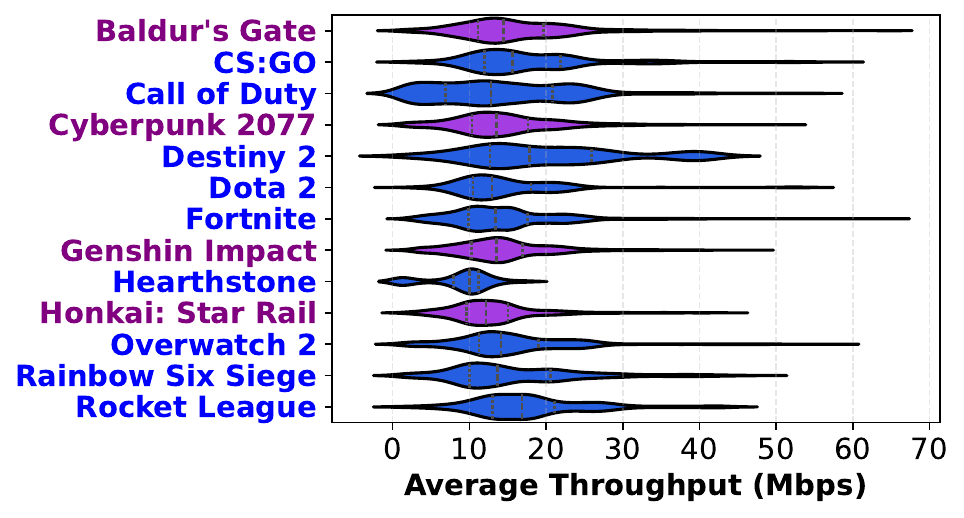}
	\label{fig:bandwidth_violin_plot_horizontal}}
    \subfigure[Average throughput per session across 2 gameplay activity patterns.]{\includegraphics[width=\columnwidth]{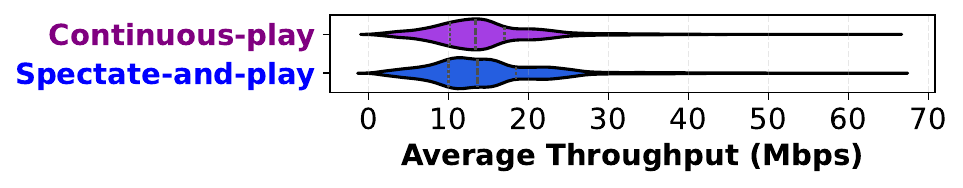}
	\label{fig:bandwidth_violin_plot_horizontal_by_type}}
    \vspace{-3mm}
    \caption{Average throughput per game streaming session for (a) 13 popular game titles and (b) 2 gameplay activity patterns.}
    \label{fig:bandwidth_violin_plot}
\end{figure}

The average bandwidth demands of streaming sessions that belong to the thirteen popular game titles and the two types of gameplay activity patterns are provided in Fig.~\ref{fig:bandwidth_violin_plot_horizontal} and Fig.~\ref{fig:bandwidth_violin_plot_horizontal_by_type} as representative examples, respectively.
According to the throughput distributions of each game title, we can observe unique ranges across different games. For example, high-demand games like Baldur's Gate and Fortnite have their maximum session-level average throughput up to 68 Mbps; while low-demand games like Hearthstone require only 20 Mbps for gameplay sessions with the best streaming settings. There are many streaming sessions with very low streaming throughput (\eg less than 1 Mbps), which likely suffer from poor user experience due to constrained network conditions with large game streaming lag mostly over 70ms from the measured objective QoE metrics. Therefore, they are excluded from our analysis of network bandwidth demands.

Depending on the game streaming settings such as graphic resolutions and user setups (\eg PC or mobile devices), we have observed several (two to four) clusters of bandwidth demands for each game title. For example, as shown in Fig.~\ref{fig:bandwidth_violin_plot_horizontal}, Destiny 2 has three clusters of bandwidth in the ranges of 8 -- 18 Mbps, 20 -- 30 Mbps, and 35 -- 47 Mbps, respectively, each of which is mapped to a group of graphic resolution settings depending on the player devices. Detection of the player devices has been done in prior work \cite{lyu_network_2024} and is not discussed in this paper.
In Fig.~\ref{fig:bandwidth_violin_plot_horizontal_by_type}, similar throughput distributions, which are dominantly in the range of 10 -- 25 Mbps with the highest bandwidth demand exceeding 65 Mbps for streaming in very high graphic resolutions, are observed for both continuous-play and spectate-and-play games, with slightly higher bandwidth demands seen for spectate-and-play games. 

With a concrete understanding of the throughput requested for streaming sessions of certain game titles, our partnered ISP is better equipped to prioritize premium users with the appropriate QoS (bandwidth) profiles for service quality assurance without over-provisioning the constrained network resources.

\begin{figure}[t!]
    \centering
    \subfigure[Objective/effective session QoE across 13 popular game titles.]{
        \includegraphics[width=\columnwidth]{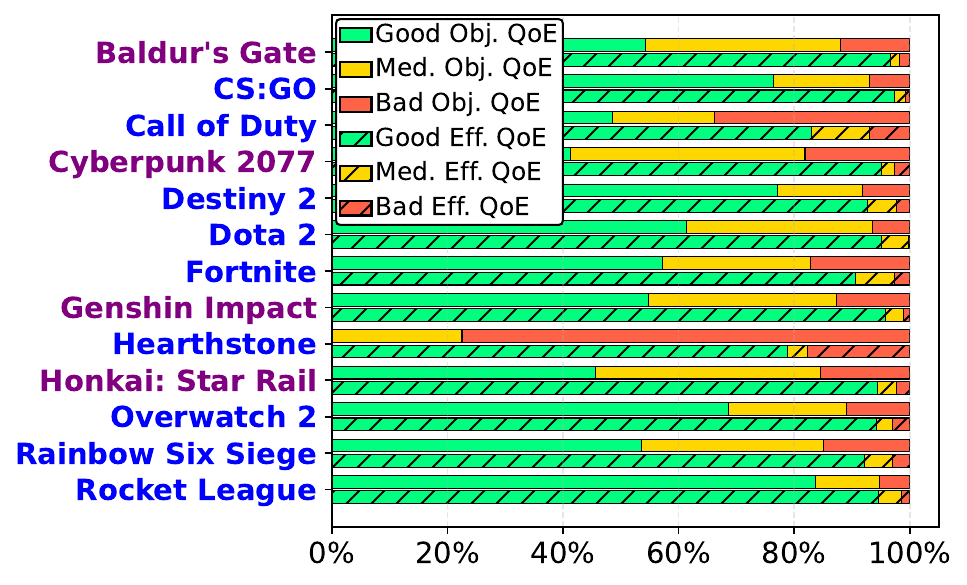}
        \label{fig:qoe_stacked_horizontal_bars_by_game}
    }
    \subfigure[Objective/effective session QoE across 2 gameplay activity patterns.]{
        \includegraphics[width=\columnwidth]{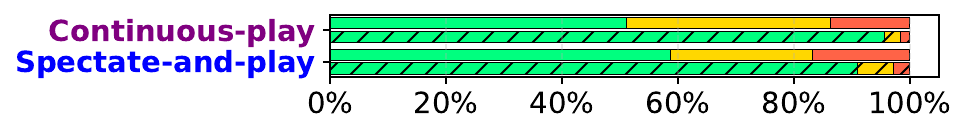}
        \label{fig:qoe_stacked_horizontal_bars_by_game_type}
    }
    \vspace{-3mm}
    \caption{Average fraction of gameplay duration per game streaming session in good, medium and bad user experience for both objective QoE and effective QoE levels, \ie before and after calibration with gameplay contexts, across (a) 13 popular game titles and (b) 2 gameplay activity pattern types.}
    \label{fig:insights_effective_qoe}
\end{figure}

\subsection{Measuring Effective Game Streaming Experience with Contexts}  \label{sec:effective_qoe}
Most importantly, by combining our measurement of cloud gaming contexts with objective QoE metrics (\eg streaming frame rate) and standard network QoS metrics (\ie latency, packet drop rate, and throughput), our partnered network operator is now able to measure the effective game streaming experience as perceived by subscribers, without mislabeling streaming sessions of certain game titles or periods of gameplay in idle/passive player activity stages that inherently require low streaming frame rate and network bandwidth as degraded user experience.

\textbf{Calibrating user experience measurement:} 
In the three-month deployment, our measurement system is operated in parallel with an existing network observability module (the bottom gray box in Fig.~\ref{fig:pipeline}) operated by our partnered ISP, which labels the objective quality-of-experience (QoE) of game streaming sessions into three levels as bad, medium or good, by mapping the measured frame rate, throughput, latency and packet drop rate of streaming flows into the expected value ranges that are internally maintained in the observability system. For example, a session with a streaming frame rate lower than 30 FPS and/or a throughput below 8 Mbps will be labeled with bad objective QoE. Augmented with the results produced in this work, the expected value ranges of the objective QoE levels are empirically calibrated according to our real-time measurement of cloud game contexts, annotated as \textbf{effective QoE levels}. After calibration, reasonable drops in \textbf{streaming frame rate} and \textbf{throughput} due to less demanding game titles (\eg Hearthstone and Honkai: Star Rail) or player activity stages (\ie idle and passive), which intrinsically have less dynamic graphics or infrequent scene changes and hence require lower network resources, will not be mislabeled with degraded QoE levels.
The expected value ranges of \textbf{low latency} and \textbf{packet drop rate} between players and cloud servers remain unchanged from the objective QoE levels to the effective QoE levels after the context-based calibration.

In Fig.~\ref{fig:insights_effective_qoe}, we show how cloud gaming contexts help our partnered ISP reduce falsely labeled sessions with degraded game streaming user experience, \ie by transitioning from objective QoE to effective QoE levels. 
For simplicity, we show the overall QoE levels of each game streaming session, representing the majority QoE labels measured for the respective session in real-time. It is clear from Fig.~\ref{fig:qoe_stacked_horizontal_bars_by_game} that all game titles have significantly more sessions with good QoE levels after being calibrated with the gameplay contexts. For example, all game streaming sessions of Hearthstone, a low-demanding card game, have medium or bad objective QoE levels, 80\% of which are corrected to the good effective QoE level. Another example is Cyberpunk 2077, which has 40\% and 16\% of its streaming sessions labeled as medium and bad for the objective QoE levels. Although this game title has high demands for network bandwidth and frame rate during the active stage, a large fraction of its player activity stages are passive and idle, leading to 95\% sessions with good effective QoE levels after calibration. 

For streaming sessions of less popular game titles that are classified as continuous-play or spectate-and-play games, in Fig.~\ref{fig:qoe_stacked_horizontal_bars_by_game_type}, we can also observe that about half of the total sessions are corrected from medium or poor in objective QoE levels to good in effective QoE levels.

By calibrating the mapping criteria of game streaming sessions with our real-time classification of cloud gameplay contexts for effective QoE measurement, our partnered network operator is able to precisely identify the groups of under-performing game streaming sessions in a much smaller (and therefore manageable) quantity, such as those from clients connected to its 5G home broadband network via a poorly configured mobile cell. It enables precise troubleshooting without mislabeling streaming sessions with low network demands as poor user experience. In addition, knowing the effective QoE after our context-based calibration helps the ISP to reactively improve under-performing game streaming sessions served by its 5G broadband network by enforcing QoS profiles for the respective sessions via network slices with higher capacity.
\section{Related Work} \label{sec:related}

\textbf{Network traffic analysis of cloud gaming services:}
As an emerging type of multimedia streaming service that exhibits high demands on network conditions, cloud gaming services have been analyzed in prior works for their network traffic characteristics, including flow anatomy and volumetric profiles. For example, Lyu \textit{et al.} \cite{lyu_network_2024} discussed the use of network flows during entire user sessions, from platform administration to server selection and gameplay, which serve as indicative signatures for network operators to detect cloud gaming sessions played on different types of user setups by their broadband/mobile subscribers. The works in \cite{lyu_do_2024,carrascosa_cloud_2022,domenico_network_2021} analyzed the bandwidth demands of cloud gaming sessions with different levels of streaming quality settings.
Furthermore, objective quality-of-experience metrics (\eg user input lag, graphic resolution, and streaming frame rates) have also been derived from network quality-of-service (QoS) attributes of game streaming flows \cite{monaco_real-time_2025,lyu_network_2024,baena_measuring_2023}.
For the network observability industry, it has been highlighted that purely using objective QoE metrics such as frame rate cannot effectively measure the quality of game streaming perceived by players, as gameplay contexts, including game titles and player activities, can significantly vary the expected/requested QoE and network QoS levels for good user experience \cite{an_tooth_2025,carvalho_qoe_2024,he_vigor_2024,wu_zgaming_2023,saha_study_2023,iqbal_dissecting_2021,lindstrom_cloud_2020}.
To bridge this gap, we develop a traffic analysis method to classify the contexts of cloud gameplay, enabling network operators to effectively measure user-perceived streaming quality when correlating with the objective QoE metrics.

\textbf{Measuring multimedia services with contexts:}
Gaining visibility into the coarse categories of multimedia streaming contexts has become important for the network observability industry to effectively measure user-perceived streaming quality as delivered by networks \cite{wang_standardizing_2024,varvello_performance_2022,bronzino_inferring_2019,barakabitze_qoe_2018,skorin-kapov_survey_2018,finamore_youtube_2011,mok_inferring_2011}. Prior works have developed methods to classify contexts in popular types of multimedia services such as video streaming \cite{wang_characterizing_2024,zhang_sensei_2021,adarsh_too_2021,bronzino_traffic_2021}, live streaming \cite{madanapalli_reclive_2021,loh_machine_2021}, video conferencing \cite{sharma_estimating_2023,michel_enabling_2022,macmillan_measuring_2021}, voice calling \cite{mauro_experimental_2020,holub_analysis_2018}, online gaming \cite{madanapalli_know_2022,schmidt_modeling_2021}, and virtual reality applications \cite{lyu_metavradar_2023,cheng_are_2022}.
For example, the works in \cite{che_packet_2012,chen_network_2006} focus on the contexts in online gaming that vary the expected network conditions for a good gameplay experience including game titles, genres, the number of players, peripheral status and team communication model.
In \cite{lyu_metavradar_2023,cheng_are_2022}, the authors measured the network demands for a good VR experience that depend on the user activity status and the number of surrounding users.
This work is specialized in classifying the contexts (\ie game titles, gameplay activity patterns, and player activity stages) of cloud game streaming sessions. As demonstrated in a large-scale deployment at an ISP hosting NVIDIA's cloud gaming servers, our method, combined with objective QoE measurement, holds unique value for network operators to better support this highly demanding service by effectively measuring the user-perceived streaming quality.

\section{Conclusion}
In this paper, we have developed a real-time network traffic analysis method that classifies the gameplay contexts of cloud game streaming sessions for network operators to effectively measure the user experience of this emerging application that exhibits high demands on network resources. We first systematically analyze unique network traffic characteristics of cloud game streaming sessions across various gameplay contexts, including packet group profiles for different game titles and flow volumetric behaviors for player activity stages. Driven by these insights, our real-time network traffic analysis method is developed with two novel processes leveraging machine learning models to classify game titles during the game launching stage and continuously measure player activity stages, respectively.
Our classification method is implemented in the network observability platform operated for an ISP hosting NVIDIA's cloud gaming servers with representative insights over a three-month period discussed. By integrating with the existing QoE and QoS metrics of cloud gaming sessions, our method helps the ISP significantly reduce the instances of poor user experience in cloud gameplay that are incorrectly labeled due to less demanding game titles and inactive player activity stages.

\section*{Acknowledgments}
We thank our anonymous shepherd and the six reviewers for their comprehensive and insightful feedback. We thank Maheesha Perera and Craig Russell from Canopus Networks
Pty Ltd, and Jeremy Hall, Matthew Kocoski and Dylan Ryan 
from Pentanet Ltd for their system engineering
and infrastructure support. 
We thank our Masters Project students Qixin Lu, Bingxian Min, Chi Zhang and Haoran Zheng, who participated in the lab dataset collection. 
This work is supported by Australian Government's National Industry PhD Program award reference number 35063 and Cooperative Research Centres Projects (CRC-P) Grant CRCPXIV000099.

\bibliographystyle{ACM-Reference-Format}
\balance
\bibliography{reference}

\appendix

\section{Ethics}\label{sec:appendix_ethics}
We have obtained ethical clearance from our university ethics board (UNSW Human Research Ethics Advisory Panel approval reference number iRECS5933) to report the measured cloud gaming session characteristics in an aggregated manner, by analyzing real-time network traffic at our partnered Internet service provider that exclusively hosts NVIDIA's GeForce NOW cloud gaming servers and offline logs.
All user identities are anonymous, and we have made no attempt to collect or reveal any personal information. Due to commercial confidentiality restrictions, we only report percentage values after aggregation instead of exact numbers from our field deployment.

\section{Data Availability}	\label{sec:data_availability}
Our ground-truth traffic traces (PCAP files) of cloud gaming sessions with their context labels (CSV files) that contain game titles, game genres, user platforms, streaming configurations and game activity stages, which are collected in our lab setup and described in \S\ref{sec:dataset}, are publicly shared with the research community via our university cloud drive. We also share a collection of supplementary preprocessing scripts in a GitHub repository. The public links to both the dataset and the scripts are available on our website \cite{lyu_dataset} under the ``Cloud Gaming'' category.

\section{Additional Evaluation Results}\label{sec:additional_eval}
In this section, we present additional evaluation results on model selection and attribute importance for game title  (\S\ref{sec:additional_title_eval}) and gameplay activity pattern (\S\ref{sec:additional_pattern_eval}) classification.

\begin{figure}[h!]
    \centering
    \mbox{
    \hspace{-2mm}
        \subfigure[RF model.]{
        \includegraphics[width=0.33\columnwidth]{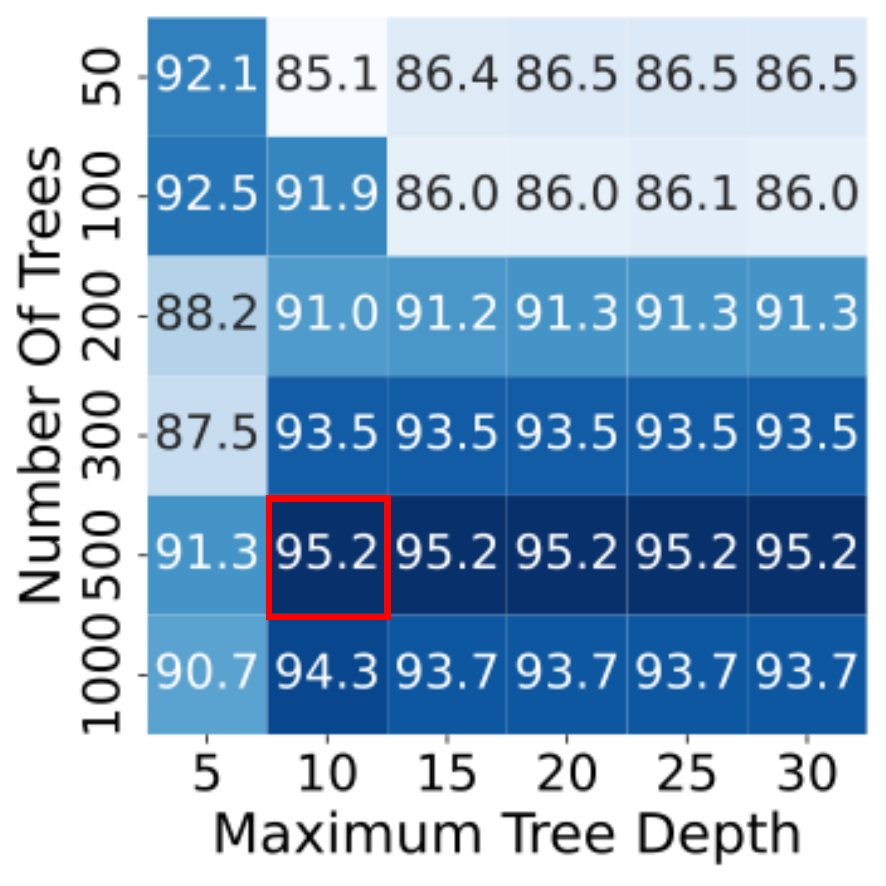}
        }
        \hspace{-3mm}
        \subfigure[SVM model.]{
        \includegraphics[width=0.33\columnwidth]{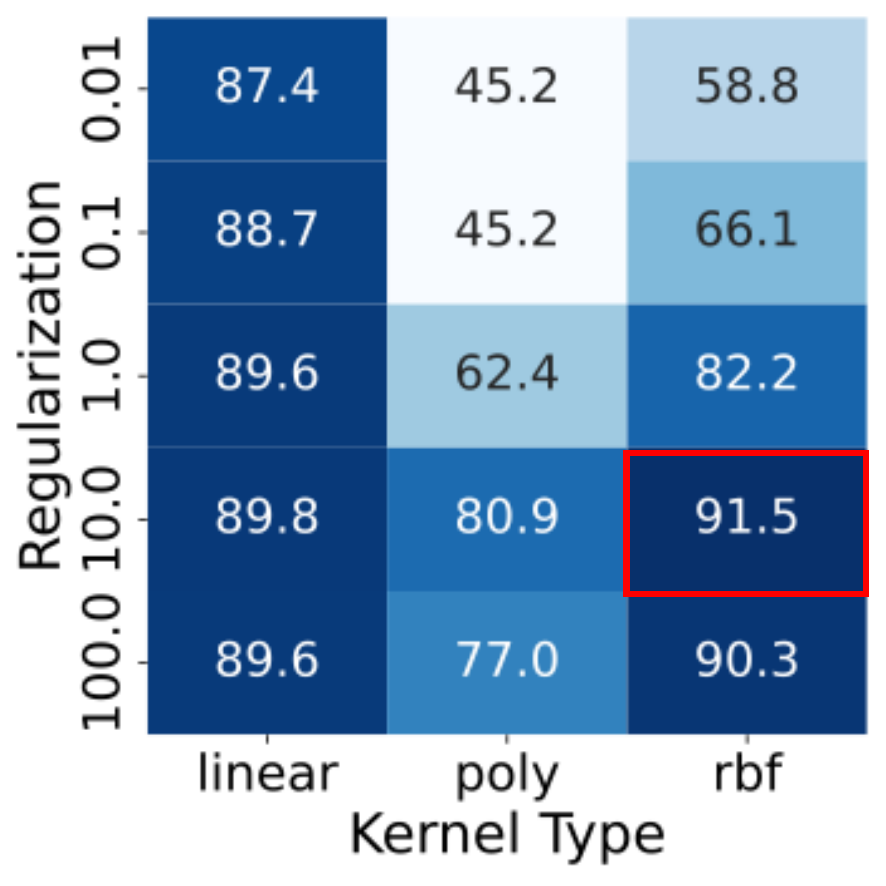}
        }
        \hspace{-3mm}
        \subfigure[KNN model.]{
        \includegraphics[width=0.33\columnwidth]{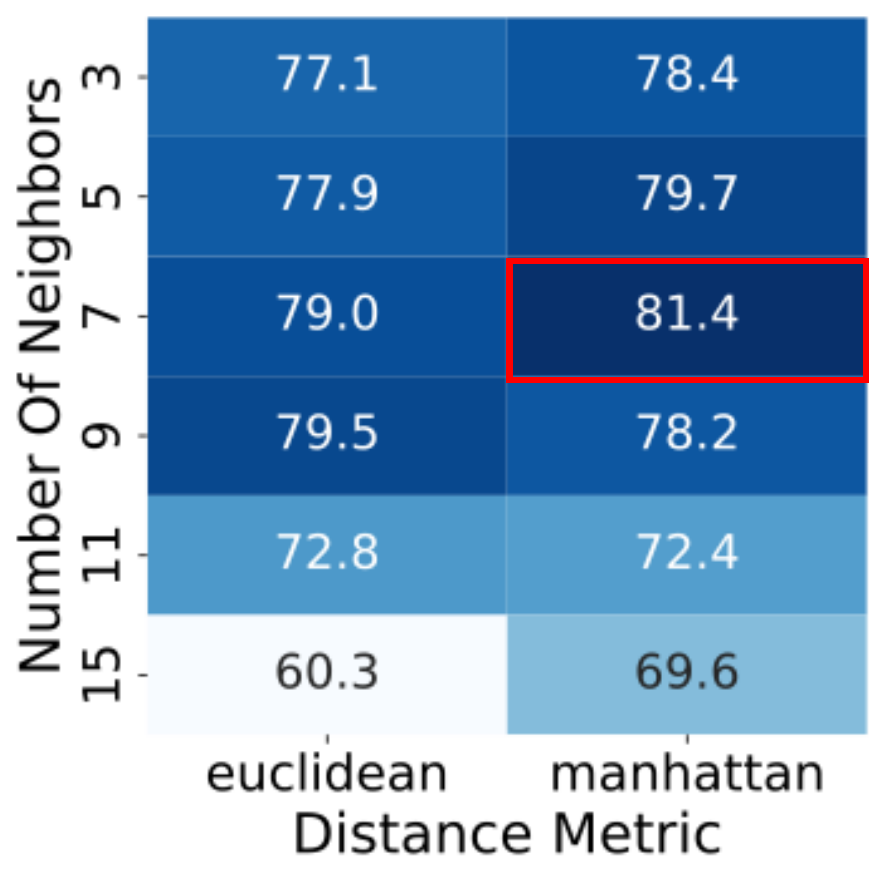}
        }
    }
    \caption{Three models (\ie RF, SVM and KNN) are fine-tuned for their hyperparameters for game title classification, with the best accuracy achieved by the RF model.}
    \label{fig:game_title_hyperparameter_tuning}
\end{figure}

\subsection{Game Title Classification}\label{sec:additional_title_eval}
We trained and fine-tuned the hyperparameters of three models that are commonly used in traffic classification tasks, namely Random Forest (RF), Support Vector Machine (SVM) and K-Nearest-Neighbors (KNN). 
In Fig.\ref{fig:game_title_hyperparameter_tuning}, we show the accuracies of the three models for game title classification while tuning their two representative hyperparameters, including the number of trees and maximum tree depth for Random Forest, regularization parameter (C) and kernel type for SVM, and the number of neighbors and distance metric for KNN. The best hyperparameter combinations from each model are highlighted in red. The overall highest accuracy (\ie 95.2\%) was achieved from the Random Forest model with the number of trees set to 500 and the maximum tree depth between 10 and 30. In our deployment, we select a maximum tree depth of 10 as it reduces model complexity as well as the risk of overfitting. In comparison, the highest accuracy from SVM and KNN models after fine-tuning are 91.5\% and 81.4\%, respectively. The substantially lower performance of the KNN model might be due to the limitations of distance-based metrics in capturing the characteristics of high-dimensional attribute space.

\begin{figure}[h!]
    \centering
    \mbox{
    \hspace{-2mm}
        \subfigure[RF model.]{
        \includegraphics[width=0.33\columnwidth]{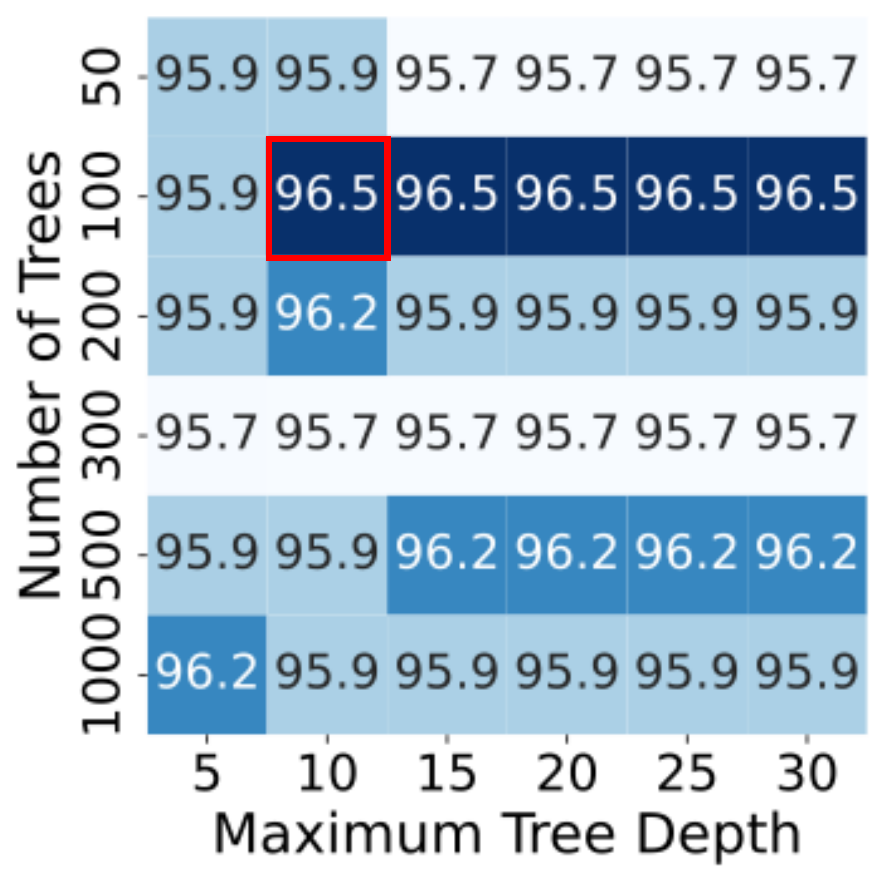}
        }
        \hspace{-3mm}
        \subfigure[SVM model.]{
        \includegraphics[width=0.33\columnwidth]{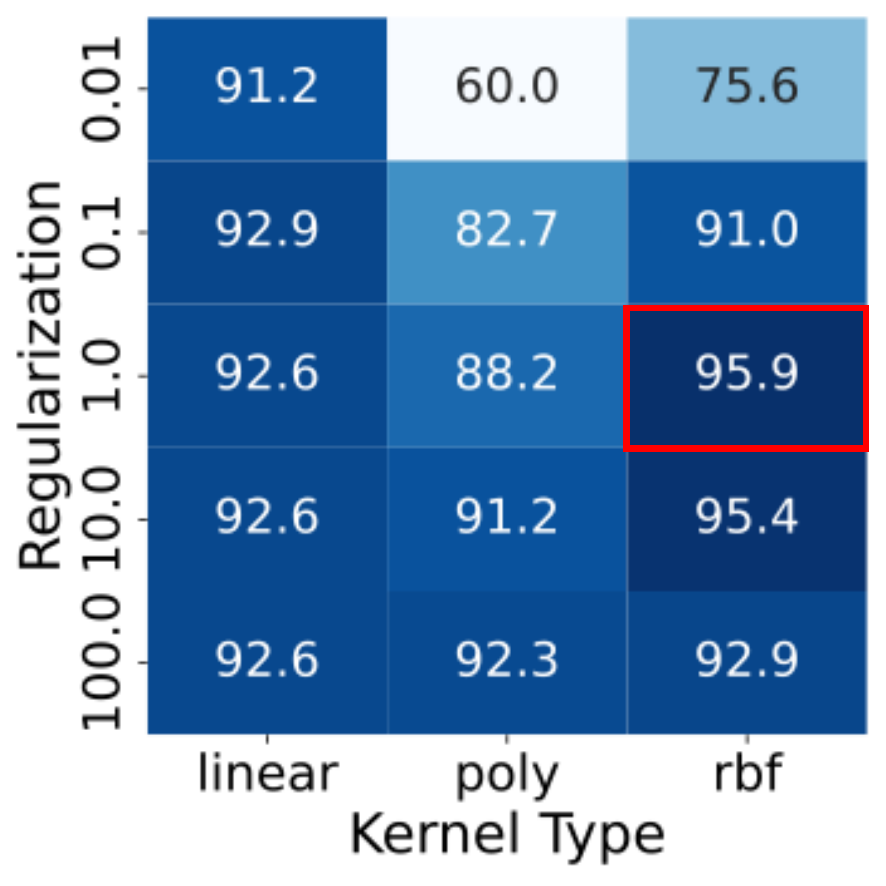}
        }
        \hspace{-3mm}
        \subfigure[KNN model.]{
        \includegraphics[width=0.33\columnwidth]{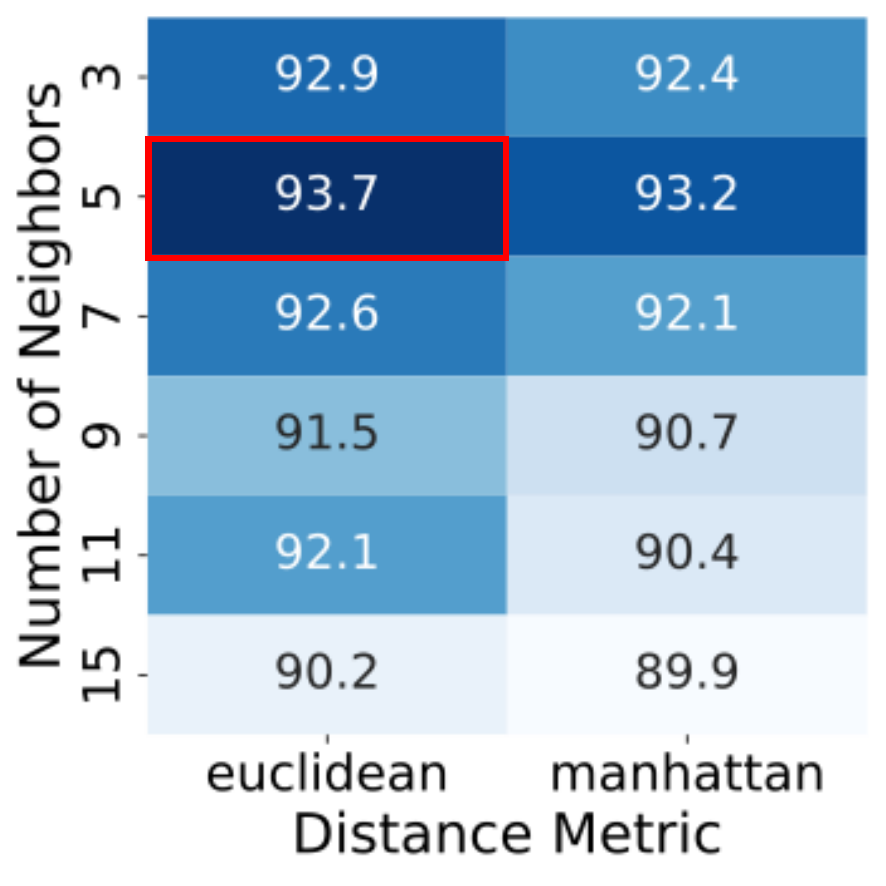}
        }
    }
    \caption{Three models (\ie RF, SVM and KNN) are fine-tuned for their hyperparameters for the classification of gameplay activity patterns, with the highest accuracy achieved by the RF model.}
    \label{fig:game_stage_hyperparameter_tuning}
\end{figure}

\subsection{Gameplay Activity Pattern Classification}   \label{sec:additional_pattern_eval}
Following the same model fine-tuning process as just described for game title classification, we show the accuracy of the three models with different hyperparameters in Fig.~\ref{fig:game_stage_hyperparameter_tuning}. Similarly, the highest accuracy of 96.5\% was achieved by the Random Forest model with the number of trees set to 100 and the maximum tree depth within the range of 10 and 30. We select a maximum tree depth of 10 in our deployment.
Due to the lower dimensionality of the attribute space, the SVM and KNN models demonstrate slightly lower accuracy compared to the Random Forest model, with their highest accuracy being 95.9\% and 93.7\%, respectively.

\begin{table}[h!]
	\centering
	\caption{Importance of the nine attributes (each representing a transition probability among the three types of player activity stages) in classifying gameplay activity patterns by the best-performing Random Forest model.}
	\small
	\begin{tabular}{|
    >{\columncolor[HTML]{C0C0C0}}l |l|l|l|}
    \hline
    \textbf{\diagbox{To}{From}}        & \cellcolor[HTML]{C0C0C0}\textbf{\coloractive{Active}} & \cellcolor[HTML]{C0C0C0}\textbf{\colorpassive{Passive}} & \cellcolor[HTML]{C0C0C0}\textbf{\coloridle{Idle}} \\ \hline
    \textbf{\coloractive{Active}}  & 0.022                                   & 0.009                                    & 0.018                                 \\ \hline
    \textbf{\colorpassive{Passive}} & 0.052                                   & 0.027                                    & 0.094                                 \\ \hline
    \textbf{\coloridle{Idle}}    & 0.167                                   & 0.022                                    & 0.026                                 \\ \hline
    \end{tabular}
	\label{tab:activity_pattern_feature_importance}
\end{table}

Table~\ref{tab:activity_pattern_feature_importance} shows the permutation importance of the nine attributes in classifying the gameplay activity pattern by our best-performing Random Forest model. Each of these attributes represents the probability of one possible stage transition.
All nine attributes have certain predictive power in classifying gameplay activity patterns. Among them, the attribute for transitions from active to idle stage has the highest importance.

\end{document}